\documentclass[final]{cvpr}
\usepackage{times}
\usepackage{epsfig}
\usepackage{graphicx}
\usepackage{amsmath}
\usepackage{amssymb}

\DeclareMathOperator{\R}{\mathbb{R}}
\newcommand{\argmax}{\arg\!\max}
\usepackage{array}
\usepackage{tabularx}
\usepackage{multirow}
\usepackage{subfigure}
\usepackage{algpseudocode}
\usepackage{color}
\usepackage{booktabs}
\usepackage{comment}

\graphicspath{{./figures/}}

\usepackage[pagebackref=true,breaklinks=true,colorlinks,bookmarks=false]{hyperref}

\def\cvprPaperID{10} 
\def\confYear{CVPR 2021}
\begin{document}
	
\title{RCNN-SliceNet: A Slice and Cluster Approach for Nuclei Centroid Detection in Three-Dimensional Fluorescence Microscopy Images}
\author{\parbox{17cm}{\centering
		{Liming Wu$^{\star}$ \quad Shuo Han$^{\star}$ \quad Alain Chen$^{\star}$ \\ Paul Salama$^{\dagger}$ \quad Kenneth W. Dunn$^{\ddagger}$  \quad Edward J. Delp$^{\star}$}\\
		{\normalsize
			$^{\star}$ Video and Image Processing Laboratory (VIPER), Purdue University, West Lafayette, Indiana\\
			$^{\dagger}$ Department of Electrical and Computer Engineering, Indiana University-Purdue University, Indianapolis, Indiana\\
			$^{\ddagger}$ Division of Nephrology, School of Medicine, Indiana University, Indianapolis, Indiana}}
}
\maketitle

\begin{abstract}
	Robust and accurate nuclei centroid detection is important for the understanding of biological structures in fluorescence microscopy images. 
	Existing automated nuclei localization methods face three main challenges: (1) Most of object detection methods work only on 2D images and are difficult to extend to 3D volumes; (2) Segmentation-based models can be used  on 3D volumes but it is computational expensive for large microscopy volumes and they have difficulty distinguishing different instances of objects; (3) Hand annotated ground truth is  limited for 3D microscopy volumes.
	To address these issues, we present a scalable approach for nuclei centroid detection of 3D microscopy volumes.
	We describe the RCNN-SliceNet to detect 2D nuclei centroids for each slice of the volume from different directions and  3D agglomerative hierarchical clustering (AHC) is used to estimate the 3D centroids of nuclei in a volume. 
	The model was trained with the synthetic microscopy data generated using Spatially Constrained Cycle-Consistent Adversarial Networks (SpCycleGAN) and tested on different types of real 3D microscopy data.
	Extensive experimental results demonstrate that our proposed method can accurately count and detect the nuclei centroids in a 3D microscopy volume.
\end{abstract}

\section{Introduction}
~\label{sec:intro}
\thispagestyle{empty}
High quality data generated by optical microscopy can provide useful information for understanding biological tissue structures ~\cite{dunn2002}.
Traditional optical microscopes have  limitations due to light scattering, leading to images with low resolution and contrast ~\cite{booth2007adaptive}.
Two-photon fluorescence microscopy allows deeper tissue imaging with near-infrared light ~\cite{benninger2013two}. 
The strongly focused subpicosecond pulses make it possible to generate high-quality three-dimensional (3D) images in living cells and deeper tissues ~\cite{Denk1990TwophotonLS}. 

Quantification results can be useful for obtaining information for subsequent analysis such as cell tracking, disease diagnosis, and new drug development ~\cite{xie2015deep}. 
Robust and accurate nuclei counting and localization are the first steps in quantifying  biological structures. 
Due to large variations of cell type, size, or microscopy modality, nuclei counting and localization remains a challenging problem especially in a 3D volume where nuclei are crowded and overlap with each other ~\cite{hayakawa2019computational}.

Segmentation approaches are popular for nuclei counting and localization. This is because segmentation can separate foreground nuclei and background structures.
Many segmentation methods use thresholding to determine binary masks of nuclei. 
In ~\cite{7904925}  a cell nuclei segmentation method that uses median filtering, Otsu's thresholding ~\cite{Otsu}, and morphological operation to segment nuclei is described. 
Otsu's thresholding can estimate a  threshold by minimizing within-class variance of the foreground and background pixels.
Similarly, ImageJ's 3D object counter, known as JACoP ~\cite{rueden2017, bolte2006guided}, uses Otsu's thresholding to segment the 3D image volumes. It then uses 3D connected components to remove small artifacts and estimates the number of objects and centroid coordinates of each object in a 3D volume. 
The counting and localization accuracy still suffers from over-segmentation caused by artifacts and noise.
The approach in ~\cite{al2009improved} was integrated in ~\cite{FARSIGHT} and proposed a graph-cuts-based binarization that used multi-scale Laplacian-of-Gaussian filtering to extract the image foreground.
CellProfiler ~\cite{CellProfiler} provides customized ``pipelines'' by adding different functional modules for image segmentation and quantitative analysis. 

Another successful method for segmentation is active contours \cite{Snakes}. It minimizes an energy function with the assumption that the approximate shape of the boundary is known. As an extension, a segmentation method known as Squassh \cite{paul2013coupling, Squassh} couples image segmentation with image restoration and uses generalized linear models for energy minimization. It enables co-localization and shape analyses for better quantifying subcelluar structures. Similarly, a 3D region-based active contour method described in \cite{soonamActiveContour} incorporated the 3D inhomogeneity correction for solving the inhomogeneous intensity issue.
These segmentation techniques cannot easily separate touching objects. 

Watershed segmentation has been used for this problem. Traditional watershed \cite{watershed} uses the regional minimum as the flood point and builds barriers when different water sources meet which results in over-segmentation. This was improved in \cite{4012364} by using a new marker-controlled watershed segmentation with conditional erosion.
Similarly, Volumetric Tissue Exploration and Analysis (VTEA) described in \cite{winfree2017quantitative} is an image analysis toolbox integrated in ImageJ which uses 2D watershed segmentation to separate different nuclei on each focal plane and merges the 2D segments to 3D segments based on the segments centroid distance. It can separate different nuclei and provide centroid information of each nucleus.

More recently,  convolutional neural networks (CNN) has been widely used in microscopy image segmentation. 
The popular image segmentation framework known as SegNet~\cite{segNet} is an encoder-decoder architecture. 
It was originally used for scene understanding applications and has been extended to 3D and used in fluorescence microscopy volume segmentation \cite{chichenCNN, DivadCVPRW}.
Similarly, \cite{VNet} and \cite{3DUnet}, an encoder-decoder type of fully convolutional neural network was proposed for dense volumetric segmentation. 
It uses deconvolution and shortcut concatenation to better reconstruct the location and context information, which result in better segmentation results in biomedical research with limited labeled data.
An improved 3D U-Net using watershed, known as  DeepSynth, was described  to segment different nuclei 
instances~\cite{deepsynth}.
The nuclei centroid detection accuracy of these methods normally depends on the watershed segmentation accuracy.

To distinguish different objects, the region proposal based network, known as regional convolutional neural network (R-CNN), was proposed in \cite{RCNN} for object localization. 
It uses selective search to generate regions of interest (RoIs) that may contain candidate objects and classify the RoIs into different categories. 
However, this network is relatively  slow for training and inference.
This has been improved in \cite{Fast-RCNN} by directly extracting the RoIs from the feature map that generated from another deep CNN.
More recently, \cite{Faster-RCNN} replaced selective search with a new region proposal network (RPN) for learning the RoIs. This significantly reduced the training time.
Alternatively, another category of object detectors \cite{liu2016ssd, redmon2018yolov3} that is based on one-stage regression network achieved competent results.
In \cite{hung2017applying}, the Faster R-CNN was used to detect nuclei in malaria images and in \cite{LimingThesis}, the modified Faster R-CNN was used for pneumonia detection in CT images. 
These object detection models only work on 2D images and are difficult to extend to 3D.

An approach described in \cite{FrustumPointNets} took advantage of 2D object detectors and estimated the 3D bounding boxes from the segmented pointcloud. But the 3D location accuracy is highly dependent on the 2D object detection accuracy. This has been improved in \cite{HoughVoting} with a 3D object detection network that is based on deep Hough voting model to directly regress the 3D object centroid. 
Similarly, \cite{sliceNet} proposed a SliceNet using the slice-and-fuse strategy for object detection in high-resolution 3D volumes. It detects the object on each slice of a volume from $3$ different directions and merges the results to a fused 3D detection.

Another challenge of deep learning-based methods is the lack of ground truth labels. 
Manually labeling ground truth masks or bounding boxes is labor-intensive and time-consuming even for an experienced person especially for 3D volumetric data. 
One way to address this problem is using data augmentation methods to create some ``new" training samples, which include some linear and non-linear transformations such as horizontal flipping, random cropping, color space transformations, or elastic deformations \cite{montserrat2017training, castro2018elastic}. 
These traditional data augmentation methods are not appropriate if the available training samples are limited. 
An alternative way to address this problem is to generate synthetic data.
In \cite{tremblay2018training}, synthetic objects are rendered with random background, lighting, and texture are used for training and testing Faster R-CNN. More recently, \cite{DivadCVPRW, SpCycleGAN, chen_2021} describe an approach for generating synthetic ellipsoidal and non-ellipsoidal nuclei using Generative Adversarial Networks (GANs) in 3D microscopy volumes.

\begin{figure*}
	\centering
	\epsfig{figure=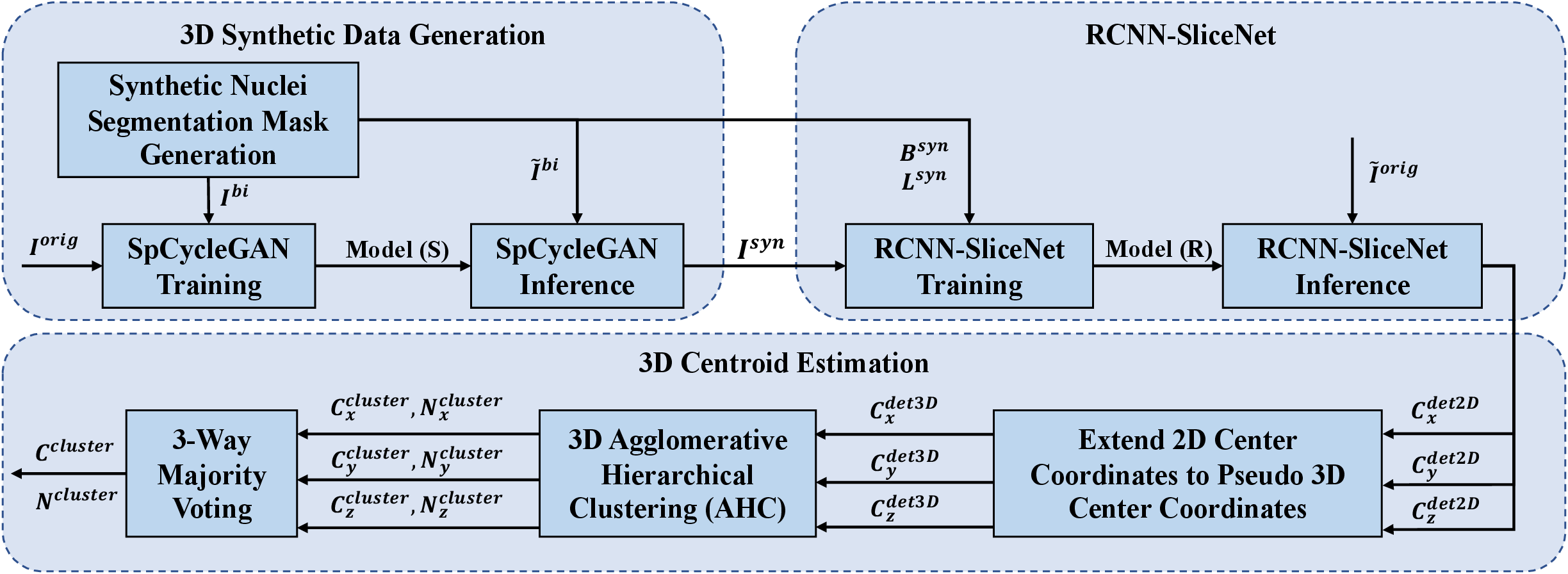,width=\textwidth}
	\caption{The block diagram of the proposed method}
	\label{fig:block_diagram}
\end{figure*}
In this paper, we present a slice-and-cluster strategy inspired by \cite{FrustumPointNets, sliceNet} that combines the Spatially Constrained CycleGAN \cite{SpCycleGAN}, RCNN-SliceNet and 3D agglomerative hierarchical clustering (AHC) to detect the centroid of nuclei in a 3D microscopy volume without the need of large amounts of ground truth labels.
We tested our method on four different real 3D microscopy data and the evaluation results show our method outperforms the rest of the baseline methods that widely used in biomedical research.

This paper provides an  approach for detecting nuclei centroids in large 3D microscopy volumes. The main contribution includes: (1) We propose the slice-and-cluster strategy for extending RCNN-SliceNet's detection results from 2D slices to 3D volumes; (2) We introduce a nuclei centroid estimation method using 3D agglomerative hierarchical clustering to estimate the true 3D nuclei centroids by making use of the RCNN-SliceNet inference results; (3) 
We provide extensive experimental results to validate the effectiveness of our designed approach. We demonstrate that our proposed method achieves best results both visually and numerically compared to other baseline models and image analysis toolkits. We also show our method is robust to generalize to other microscopy data without training.

\section{Proposed Method}
\label{sec:method}
As shown in Figure \ref{fig:block_diagram}, the block diagram of the proposed system consists of three major parts, 3D synthetic data generation, RCNN-SliceNet, and 3D centroid estimation. 

In this paper, $I$ denotes a 3D image volume of size $X\times Y\times Z$. $I_z$ denotes all slices on the $z$-direction, and $I_{z_p}$ is denoted as the $p^{th}$ slice, which is of size $X\times Y$, along the $z$-direction in a volume, where $p\in \{1,...,Z\}$. Similarly, $I_{x_p}$ and $I_{y_p}$ denote the $p^{th}$ focal plane image along the $x$-direction and along the $y$-direction.
We then denote $I^{orig}$ as the original microscopy volume of size $X\times Y\times Z$. Similarly, $I^{bi}$ denotes the synthetic binary volume and $I^{syn}$ denotes the synthetic microscopy volume. For example, $I^{orig}_{z_{25}}$ is the $25^{th}$ focal plane of the original microscopy volume along the $z$-direction. 
$I^{bi}$ and $\tilde{I}^{bi}$ are different binary volumes use for training and inference of SpCycleGAN~\cite{SpCycleGAN}. Similarly, $I^{orig}$ and $\tilde{I}^{orig}$ are different microscopy volumes.

$B^{syn}$ is a $N$ by $4$ matrix that represents the 2D coordinates of the bounding box for the nuclei on each slices along a direction.
$L^{syn}$ is the label of each bounding box, in our case, all $L^{syn}=1$ which indicates the nucleus.
Finally, as shown in Figure \ref{fig:block_diagram}, the trained RCNN-SliceNet is used to inference on the $x-$, $y-$ and $z-$ directions of a volume.
Suppose RCNN-SliceNet is used on the $z$-direction then we denote $C^{det2D}_{z}$ as the detected centroid coordinates of nuclei in all 2D slices along the $z$-direction of a volume, and $C^{det3D}_{z}$ is the pseudo 3D detected centroid coordinates estimated from $C^{det2D}_{z}$.

$C^{cluster}_z$ is the 3D coordinates of cluster centroids and $N^{cluster}_z$ is the number of clusters. Similarly, if the RCNN-SliceNet is used on the $x$-direction and the $y$-direction, $C^{det2D}_{x}, C^{det3D}_{x}, C^{cluster}_{x}, N^{cluster}_{x}$, and $C^{det2D}_{y}, C^{det3D}_{y}, C^{cluster}_{y}, N^{cluster}_{y}$ will be generated.
\begin{figure*}
	\centering
	\epsfig{figure=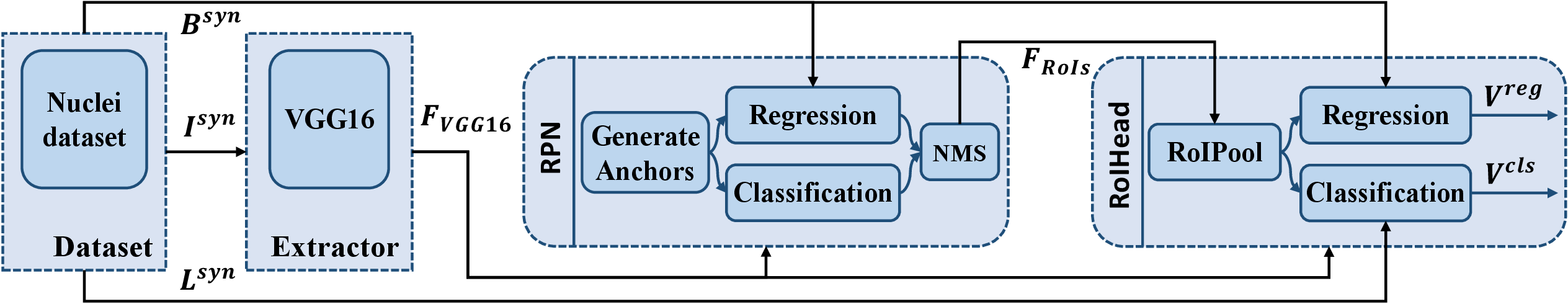,width=\textwidth}
	\caption{Block diagram of the RCNN-SliceNet}
	\label{fig:rcnn}
\end{figure*}
\subsection{3D Synthetic Data Generation}
\label{ssec:synthetic_data}
Synthetic data generation consists of synthetic nuclei segmentation mask generation and synthetic microscopy volume generation (Figure \ref{fig:block_diagram}).

The nuclei are assumed to have ellipsoidal shape based on the observation of the original microscopy volume.
Candidate nuclei $I^{can}$ in different size, location and orientation are generated in a 3D volume $I^{bi}$. The affine transformation described in \cite{DivadCVPRW} is used to generate nuclei in different orientation. The size of $I^{can}$ is determined by the length of semi-axes $\textbf{a}=(a_x,a_y,a_z)$, which are appropriately chosen based on the nuclei size in original microscopy volume. The $k^{th}$ nucleus candidate $I^{can,k}$ with intensity $k$ is generated by Equation \ref{eq:syn_bi}. 
In order to differentiate nuclei instances, each candidate nucleus $I^{can,k}$ has a unique gray-level intensity $k$. 
	The maximum overlap between two candidate nuclei is set to be at a threshold of $T_{ov}$ voxels.
\begin{align}\label{eq:syn_bi}
	I^{can,k}=
	\begin{cases}
		k, & \text{if  } \frac{\tilde{x}^2}{a_x}+\frac{\tilde{y}^2}{a_y}+\frac{\tilde{z}^2}{a_z}<1\\
		0,& \text{otherwise}
	\end{cases}
\end{align}

Candidate nucleus $I^{can,k}$ is added to $I^{label}$, which is initialized as a $128\times128\times128$ volume with intensity $0$. 
$B^{syn,k}$ and $L^{syn}$ are the ground truth bounding boxes coordinates and the category of the $k^{th}$ nucleus.

The spatially constrained CycleGAN (SpCycleGAN) from  \cite{SpCycleGAN} is used for synthetic microscopy volume generation. SpCycleGAN is an extension of CycleGAN \cite{CycleGAN} with an additional generative network $H$ and a spatial constraint loss $ \mathcal{L}_{spatial}$ that can generate synthetic microscopy images while maintaining the nuclei location and contour from the synthetic binary volume. 

SpCycleGAN consists of $5$ networks $G$, $F$, $H$, $D_1$, and $D_2$. $G$ maps $I^{bi}$ to $I^{orig}$, and $D_2$ attempts  to distinguish between the generated images  $G(I^{bi})$ and original images $I^{orig}$. Similarly, $F$ maps $I^{orig}$ to $I^{bi}$, and $D_1$ attempts to discriminate the binary images $I^{bi}$ and translated images $F(I^{orig})$. $G(I^{bi})$ is the translated microscopy like volume and $F(I^{orig})$ is the translated binary like volume.
To keep the spatial information consistent, the network $H$ with the same architecture of $G$ is introduced to measure the spatial constraint loss $\mathcal{L}_{spatial}$ between $F(G(I^{bi}))$ and $I^{bi}$.
The loss function of SpCycleGAN is described in Equation \ref{eq:spcyclegan_loss}
\begin{align}\label{eq:spcyclegan_loss}
	\nonumber
	\mathcal{L}(G, F, H, D_2,D_1)&=\mathcal{L}_{GAN}(G,D_2, I^{bi},I^{orig})\\ \nonumber
	&+\mathcal{L}_{GAN}(F,D_1,I^{orig}, I^{bi})\\ \nonumber
	&+\lambda_1 \mathcal{L}_{cycle}(G,F,I^{orig}, I^{bi})\\
	&+ \lambda_2 \mathcal{L}_{spatial}(G,H,I^{orig}, I^{bi})
\end{align}
where $\mathcal{L}_{GAN}$ and $\mathcal{L}_{cycle}$ are the adversarial and cycle losses~\cite{CycleGAN}, and $\mathcal{L}_{spatial}$ is the spatial consistency loss \cite{SpCycleGAN}.
\subsection{RCNN-SliceNet}
\label{ssec:rcnn}
As shown in Figure \ref{fig:rcnn}, the block diagram of the proposed RCNN-SliceNet nuclei localization system can be divided into three parts: training data preparation, region proposal network, and RoIHead network.

The RCNN-SliceNet was modified based on the use of Faster R-CNN model  for pneumonia detection on CT images ~\cite{Faster-RCNN, LimingThesis}. The $13$ convolution layers from a  pre-trained VGG16 on ImageNet were used to obtain the features from the input image.
The  feature map $F_{VGG}$ is of size $(C, H, W)$, where $C$ is the number of channels and $H$ and $W$ represent the height and width of the feature map. $F_{VGG}$ and ground truth bounding boxes $B^{syn}$ are sent into a region proposal network (RPN). 
The RPN will provide the regions of interest (RoIs) $F_{RoIs}$ from $F_{VGG}$ indicating where might be an object. 
During training RPN, a sliding window moves on $F_{VGG}$, generating $9$ fixed size anchor boxes $a_{1-9}$ at each sliding position in three scales $(s_1^2,s_2^2,s_3^2)$ and three aspect ratios $(r_1,r_2,r_3)$.
$s,r$ are the area, and the ratio of width to height of the anchor $a_i$.
%
The training loss of RPN $\mathcal{L}_{RPN}$ consists of the bounding box classification loss $\mathcal{L}_{cls}$ and the bounding box regression loss $\mathcal{L}_{reg}$.
\begin{align}\label{eq:rcnn_loss}
	\mathcal{L}_{RPN}&=\mathcal{L}_{cls} + \lambda \mathcal{L}_{reg}
\end{align}
where $\lambda$ is a constant controlling the balance of $\mathcal{L}_{cls}$ and $\mathcal{L}_{reg}$~\cite{Faster-RCNN}. $\mathcal{L}_{cls}$ is measured by the negative log softmax function over $K=2$ categories.
Instead of directly regressing the bounding box coordinates, $B^{syn}$ was parameterized for better convergence. The smooth $L1$ loss function defined in \cite{Fast-RCNN} was used to measure the bounding box regression loss $\mathcal{L}_{reg}$. Non-maximum Suppression (NMS) was used to remove duplicate boxes.

As shown in Figure \ref{fig:rcnn}, the RoIs $F_{RoIs}$ along with $B^{syn}$ are sent to RoIHead network to further estimate the location and category of the nuclei. The RoIPool layer \cite{SPPNet, Fast-RCNN} will convert all $F_{RoIs}$ in different dimension to the fixed size feature maps $F_{pooled}$ and $F_{pooled}$ is mapped to two output vectors: softmax probability vector $V_{cls}\in \R^2$ and bounding box coordinates vectors $V_{reg}\in \R^{2\times 4}$. The loss function of RoIHead $\mathcal{L}_{RoIHead}$ is the same as $\mathcal{L}_{RPN}$ described in Equation \ref{eq:rcnn_loss}. 

The RCNN-SliceNet was trained on 2D slices of synthetic microscopy images $I^{syn}$ on the $z$-direction and tested on original microscopy images $I^{orig}$ along the $x$-direction, the $y$-direction, and the $z$-direction, respectively. The 2D detected centroid coordinates are denoted as $C^{det2D}_{x}$, $C^{det2D}_{y}$, and $C^{det2D}_{z}$.
\subsection{3D Centroid Estimation}
\label{ssec:centroid_estimation}
\begin{figure}[htb]
	\vspace{-0.1in}
	\centering
	\epsfig{figure=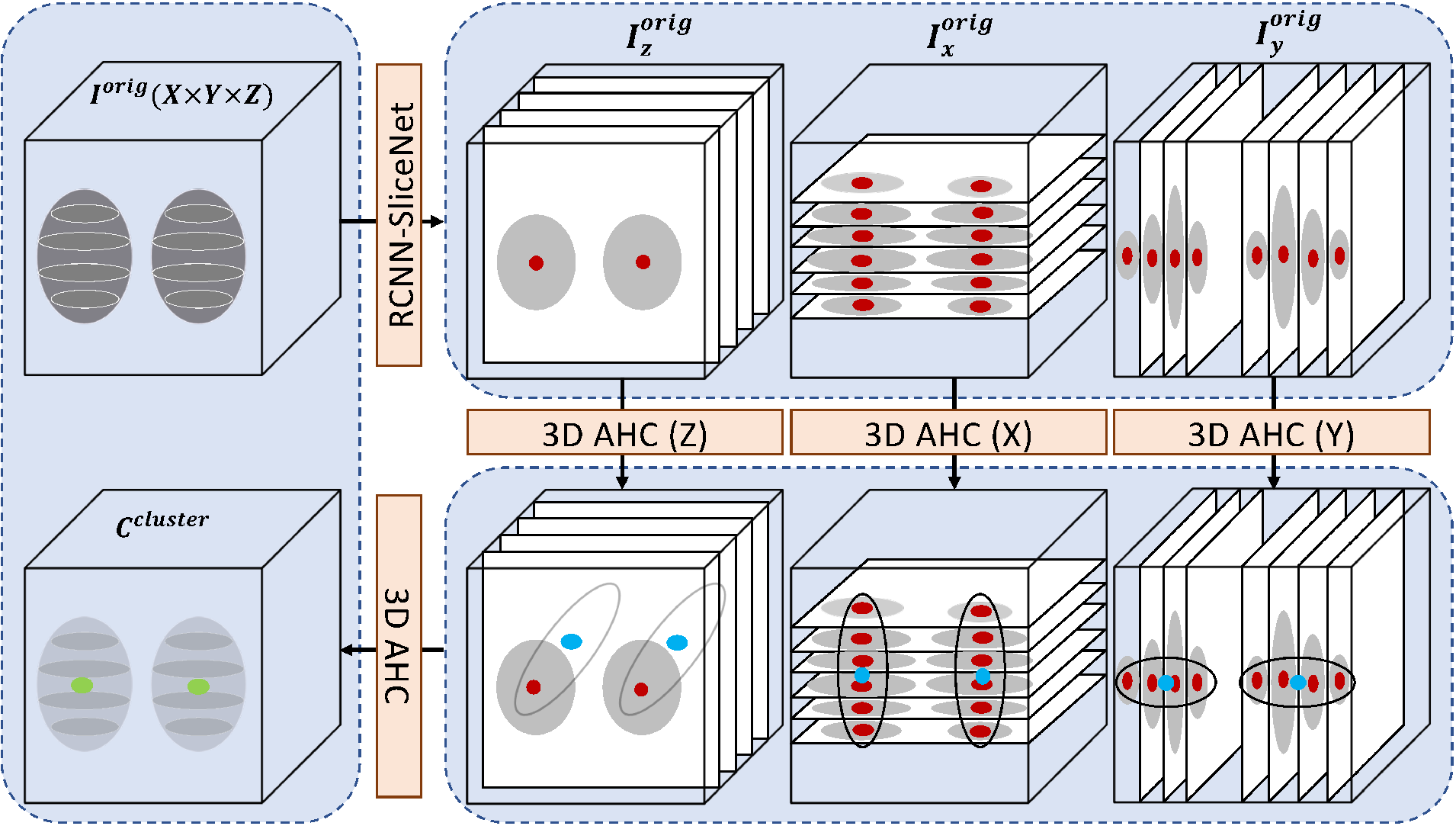,width=8.2cm}
	\caption{Overview of the RCNN-SliceNet detection and 3D AHC clustering for nuclei centroid estimation}
	\label{fig:2Dto3D}
\end{figure}

Suppose the RCNN-SliceNet is used on the $z$-direction. The detected nuclei centroids $C^{det2D}_{z}$ from model $\textbf{R}$ are 2D coordinates. For any nucleus in $I^{orig}$, multiple detected centroids will be generated in $C^{det2D}_{z}$ due to the appearance of the nucleus in multiple focal planes. Thus, as shown in Figure \ref{fig:block_diagram}, the first step of 3D centroid estimation is to extend 2D detected centroids coordinates $C^{det2D}_{z}$ to pseudo 3D detected centroid coordinates $C^{det3D}_{z}$. This step is achieved by adding the slice number as coordinates. The overview of 3D AHC clustering is shown in Figure \ref{fig:2Dto3D}. For example, if the RCNN-SliceNet inference is along the $z$-direction of $I^{orig}$ and there are $7$ nuclei were detected on $15^{th}$ focal plane. The detected 2D centroid coordinates are $C^{det2D}_{z15}$, and the detected pseudo-3D centroid coordinates $C^{det3D}_{z15}$ is generated by adding $15$ as the $z$ coordinates. Similarly, if the RCNN-SliceNet inference is along the $x$-direction of $I^{orig}$, the detected 2D centroid coordinates on the $p^{th}$ focal plane is extended to the pseudo 3D centroid coordinates $C^{det3D}_{x_p}$ by adding $p$ as the $x$ coordinates.
To estimate the 3D nuclei centroid, the agglomerative hierarchical clustering (AHC) ~\cite{HCA} with average linkage criterion is used to obtain structural information from $C^{det3D}_z$. AHC tries to create nested clusters by greedily merging a pair of clusters that have a similar property. The average linkage criterion forces AHC to minimize the average distances between all cluster pairs. The Lance-Williams dissimilarity \cite{Ward} $L$ in Equation \ref{eq:LW_dissimilarity} is used to estimate the dissimilarity between the newly merged cluster $i\cup j$ and an external cluster $e$. AHC will initially treat each point as a cluster and iteratively estimate all inter-points dissimilarities and form clusters from two closest points. Eventually all points will be merged as one cluster.
\begin{align}\label{eq:LW_dissimilarity}
	\nonumber
	L(i\cup j,e)&=\alpha_i L(i,e)+\alpha_j L(j,e)+\beta L(i, j)\\
	&+\gamma |L(i,e)-L(j,e)|
\end{align}
where $\alpha_i$, $\alpha_j$, $\beta$, $\gamma$ define the agglomerative criterion, and $|\cdot|$ represents the absolute value.

The  number of clusters $N^{cluster}_z$ is estimated using the Silhouette Coefficient \cite{silhouette}. For a given number of clusters $k, k\in [1,n]$, the mean intracluster distance $a(i)$ and the average of the mean intercluster distance $b(i)$ for each point $i \in C^{det3D}_z$ is estimated using Equation \ref{eq:intro_inter_distance}.
\begin{align}\label{eq:intro_inter_distance}
	\nonumber
	a(i)&=\frac{1}{n_c-1}\sum_{i,j\in C_c,i\neq j}{d(i,j)}\\
	b(i)&=\frac{1}{k-1}\sum_{q,q\neq c}{(\frac{1}{n_q}\sum_{i\in C_c,j\in C_q}{d(i,j)})}
\end{align}
$a(i)$ measures the distance between $i$ and all other samples in its cluster. $c$ denotes the cluster where $i$ is in, $n_c$ is the number of points in cluster $c$, $C_c$ are all points in cluster $c$, and $d(i,j)$ is the Euclidean distance between $i$ and $j$.
$b(i)$ measures the distance between $i$ and other clusters $q$. 
\begin{align}\label{eq:SC_score}
	\nonumber
	SC_k&=\frac{1}{n}\sum_{i\in C^{det3D}_z}{\frac{b(i)-a(i)}{\text{max}(a(i),b(i))}}\\
	N^{cluster}_z &= \argmax_k SC_k, k=1,...,n
\end{align}
The Silhouette Coefficient $SC_k$ when cluster number is $k$ is given by Equation \ref{eq:SC_score}. The optimal number of clusters $N^{cluster}_z$ is the number of clusters that corresponds to the highest Silhouette Coefficient, and the cluster centroid $C^{cluster}_z$ are the centroid coordinates of all clusters. Similarly, the $C^{cluster}_x$, $N^{cluster}_x$, and $C^{cluster}_y$, $N^{cluster}_y$ are obtained by repeating the entire process on the $x$-direction and the $y$-direction of the volume.

In the final step, 3-way majority voting was used to better estimate the cluster centroid. Three-way majority voting is achieved by using another 3D AHC that cluster the points of $C^{cluster}_x$, $C^{cluster}_y$, and $C^{cluster}_z$. 
Ultimately, we use $C^{cluster}$ as the centroid coordinates of the nuclei and use $N^{cluster}$ as the number of nuclei in a 3D volume.
\section{Experimental Results}
\label{sec:results}
\subsection{Datasets}
\label{ssec:experimental_setup}
We used four different types of microscopy data for evaluation, denoted as Data-I, Data-II, Data-III and Data-IV.
The Data-I and Data-II are fluorescent-labeled (Hoechst 33342 stain) nuclei collected using two-photon fluorescent microscopy of rat kidneys.
Data-I consists of one volume in size of $X\times Y\times Z=128\times 128\times 64$ pixels, Data-II consists of 16 volumes in size of $128\times 128\times 32$ pixels, Data-III consists of 21 volumes in size of $128\times 128\times 19$ pixels, and Data-IV consists of 90 volumes in dimension of $807\times 565\times 129$. Data-I, Data-II and Data-III were provided by Malgorzata Kamocka and Kenneth W. Dunn of Indiana University and were collected at the Indiana Center for Biological Microscopy \cite{indiana_center}. Data-IV is BBBC024vl \cite{svoboda2009generation} from the Broad Bioimage Benchmark Collection.

During the synthetic binary volume generation, 54 synthetic binary volumes ($I^{bi01}-I^{bi54}$) for Data-I, Data-II and Data-III are generated, respectively. 
The semi-axes \textbf{a} are randomly chosen in the range of 4 and 8 for synthetic Data-I, 10 and 14 for synthetic Data-II, 6 and 10 for synthetic Data-III. The corresponding numbers of candidate nuclei $N^{syn}$ in a volume are set to 400, 40, and 50, and the allowed overlapping threshold $T_{ov}$ are set to 5, 10, and 10, respectively.
$I^{bi001}-I^{bi004}$ along with the corresponding original microscopy data $I^{orig}$ are used to train the SpCycleGAN. The rest of synthetic binary volumes $I^{bi005}-I^{bi054}$ are inferenced on the trained SpCycleGAN to generate synthetic microscopy data $I^{syn005}-I^{syn054}$. All synthetic binary volumes $I^{bi}$ and synthetic microscopy volumes $I^{syn}$ are in size $X\times Y\times Z = 128\times128\times128$. 

\begin{figure}[htb!]
	\centering
	\vspace{-0.1in}
	\subfigure[]
	{
		\epsfig{figure=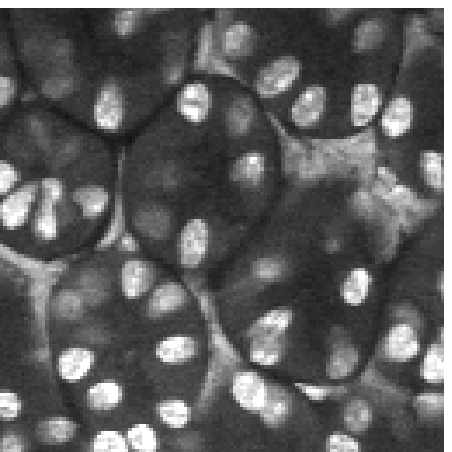,width=2.2cm}
	}
	\subfigure[]
	{
		\epsfig{figure=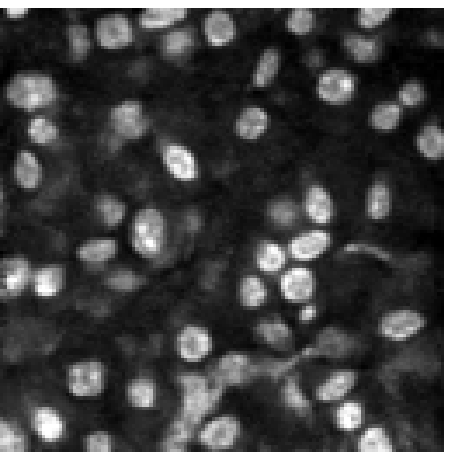,width=2.2cm} 
	}
	\subfigure[]
	{
		\epsfig{figure=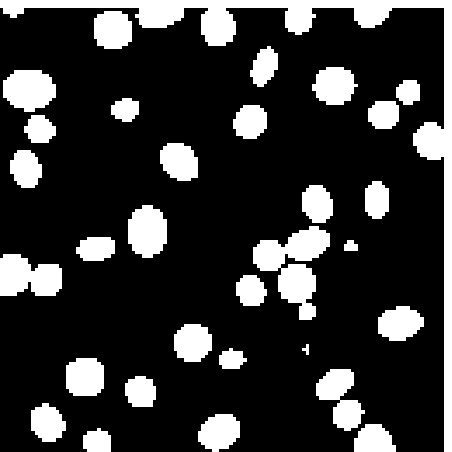,width=2.2cm}
	}
	\subfigure[]
	{
		\epsfig{figure=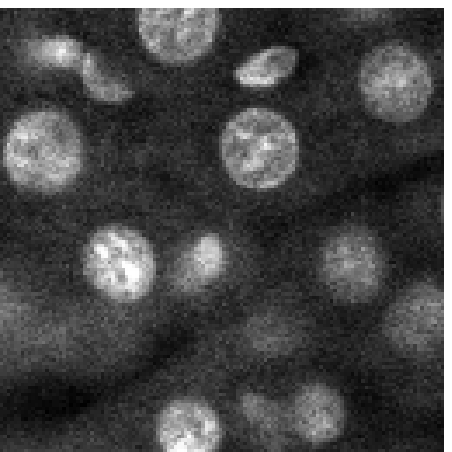,width=2.2cm}
	}
	\subfigure[]
	{
		\epsfig{figure=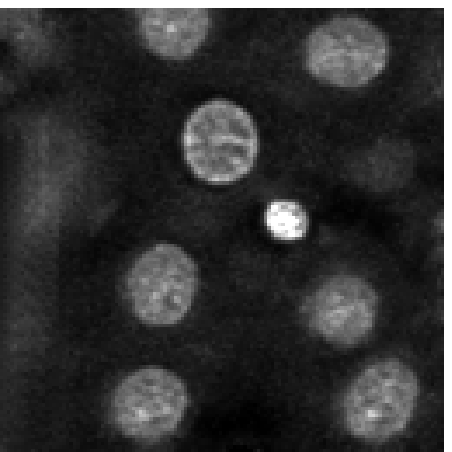,width=2.2cm} 
	}
	\subfigure[]
	{
		\epsfig{figure=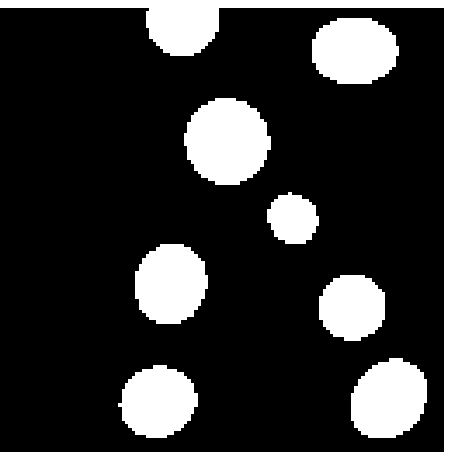,width=2.2cm}
	}
	\subfigure[]
	{
		\epsfig{figure=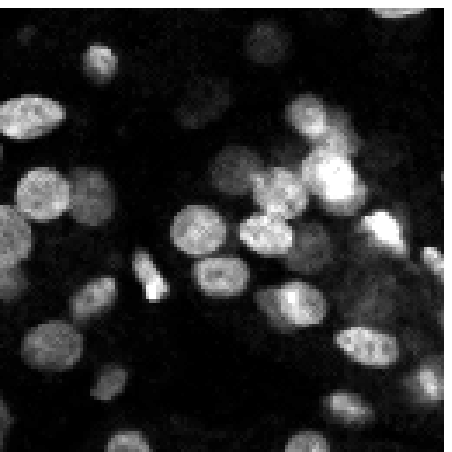,width=2.2cm}
	}
	\subfigure[]
	{
		\epsfig{figure=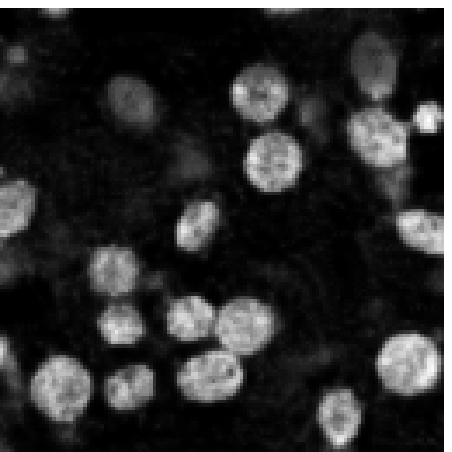,width=2.2cm} 
	}
	\subfigure[]
	{
		\epsfig{figure=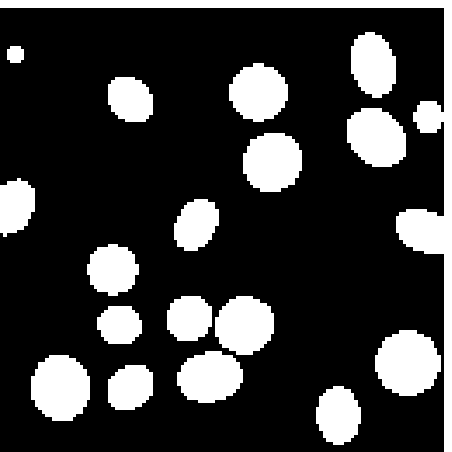,width=2.2cm}
	} 
	\caption{Synthetic microscopy images (second column) generated based on the original microscopy images (first column) and the synthetic ground truth masks (third column) using SpCycleGAN. First row is Data-I, second row is Data-II, and third row is Data-III}
	\label{fig:synimages}
	\vspace{-0.1in}
\end{figure}

Both the SpCycleGAN and RCNN-SliceNet were implemented using PyTorch. The SpCycleGAN was trained using Adam optimizer with a constant learning rate $0.0002$ for the first $100$ epochs and linearly decayed to $0$ in the next $100$ epochs. 
The generators $G, F,$ and $H$ use 9 blocks of ResNet. The loss parameters are set to $\lambda_1 = \lambda_2 = 10$. 
As shown in Figure \ref{fig:synimages}, the SpCycleGAN generates synthetic microscopy images that look like the original microscopy images. It can maintain the size and the shape of the nuclei from the binary mask.
\begin{table*}[]
	\centering
	\setlength\tabcolsep{0.07 in}
	\caption{Evaluation of nuclei counting and centroid detection on original Data-I, Data-II and Data-III. The counting accuracy is evaluated using the mean absolute percentage error (MAPE\%). The centroid-based accuracy is evaluated using average precision (AP\%) and mean Average precision (mAP\%)}
	\begin{tabular}{ccccc|cccc|cccc}
		\hline
		\multirow{2}{*}{} & \multicolumn{4}{c|}{Microscopy Data-I} & \multicolumn{4}{c|}{Microscopy Data-II} & \multicolumn{4}{c}{Microscopy Data-III} \\ \cline{2-13} 
		& MAPE & $\mathrm{AP}_4$ & $\mathrm{AP}_8$ & mAP & MAPE& $\mathrm{AP}_6$ & $\mathrm{AP}_{12}$ & mAP &MAPE & $\mathrm{AP}_{6}$ & $\mathrm{AP}_{10}$&mAP\\ \hline
		ImageJ\cite{bolte2006guided} & 43.66  & 10.86  & 21.69  & 16.48  & 16.67  & 41.88  & 57.23  & 50.22 &40.25&35.43&41.32&38.68 \\ 
		3D Watershed & 14.08  & 55.25  & 70.66  & 65.75  & 44.34  & 58.40&	65.04&	63.38 &43.02&35.91&	41.83	&39.54\\
		Squassh\cite{Squassh} & 76.06  & 9.97  & 14.38  & 12.67  & 17.63  & 44.05  & 55.13  & 50.62 &37.91&33.76&42.38&38.57 \\ 
		CellProfiler\cite{CellProfiler} & 5.28  & 41.39  & 56.95  & 51.66  & 28.40  & 62.46  & 66.70  & 65.55 &32.74&44.00&58.02&51.73 \\
		VTEA\cite{winfree2017quantitative} & 13.73  & 30.20  & 42.68  & 37.84  & 14.06  & 60.65  & 63.14  & 61.77&11.00& 63.02&69.72&66.87 \\ 
		V-Net\cite{VNet}&17.61&73.78&80.70&	78.97&27.16&49.87&56.41&54.26&11.12&61.13&69.92&66.87\\
		3D U-Net\cite{3DUnet} & 20.77&73.60&78.12&76.95&15.57&57.74&63.74&62.16&12.77&60.58&70.68&66.50\\
		DeepSynth\cite{deepsynth} & 8.80 &74.24 &81.98 &80.00 & 14.49  & 64.06&	69.45&	68.12 & 17.10&61.26&70.24&66.86\\ \hline
		Proposed & \textbf{1.41}  & \textbf{86.41}  & \textbf{87.86}  & \textbf{87.52}  & \textbf{9.76}  & \textbf{75.63}  &\textbf{76.42}  & \textbf{76.11} & \textbf{8.54} &\textbf{73.91}&\textbf{80.97}&\textbf{78.50} \\ \hline
	\end{tabular}
	\vspace{-0.1in}
	\label{table_1}
\end{table*}
\begin{table}[h!]
	\centering
	\setlength\tabcolsep{0.09 in}
	\caption{Evaluation results of microscopy Data-IV using pre-trained models on synthetic microscopy Data-III}
	\begin{tabular}{ccccc}
		\hline
		\multirow{2}{*}{} & \multicolumn{4}{c}{Microscopy Data-IV} \\ \cline{2-5} 
		&MAPE & $\mathrm{AP}_{15}$ & $\mathrm{AP}_{25}$&mAP\\ \hline
		V-Net\cite{VNet}&1.50&95.41&	95.67&	95.55\\
		3D U-Net\cite{3DUnet} & 2.00&94.30&	94.40&	94.34\\
		DeepSynth\cite{deepsynth} &2.78& 91.31&	91.94&	91.61\\ \hline
		Proposed &\textbf{1.11}&\textbf{98.71}&\textbf{98.80}&\textbf{98.79}\\ \hline
	\end{tabular}
	\vspace{0.1in}
	\label{table_2}
	\vspace{-0.2 in}
\end{table}
The RCNN-SliceNet was trained on synthetic volumes and tested on original microscopy volumes. 
The 3D ground truth masks for the original microscopy volumes are manually annotated using ITK-SNAP~\cite{ITK-SNAP}.
\begin{figure}[htb!]
	\centering
	\vspace{-0.12in}
	\subfigure[]
	{
		\epsfig{figure=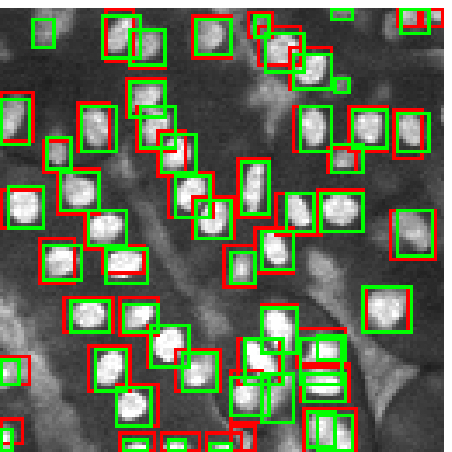,height=   2.35cm}
	}
	\hspace{-0.3cm}
	\subfigure[]
	{
		\epsfig{figure=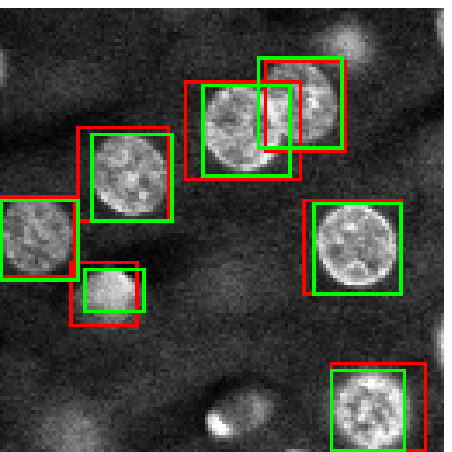,height=    2.35cm}
	}
	\hspace{-0.3cm}
	\subfigure[]
	{
		\epsfig{figure=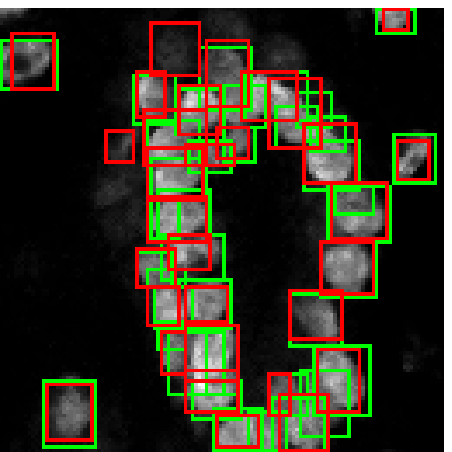,height=    2.35cm}
	}
	\hspace{-0.3cm}
	\subfigure[]
	{
		\epsfig{figure=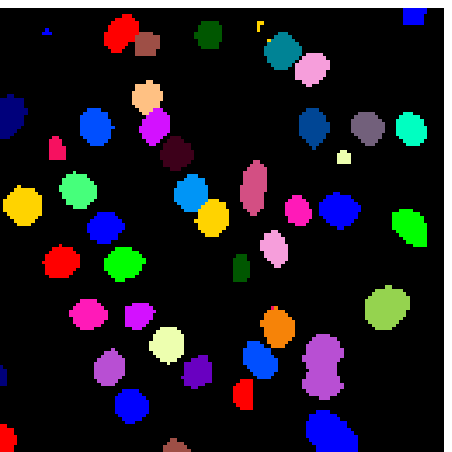,height=    2.35cm} 
	}
	\hspace{-0.3cm}
	\subfigure[]
	{
		\epsfig{figure=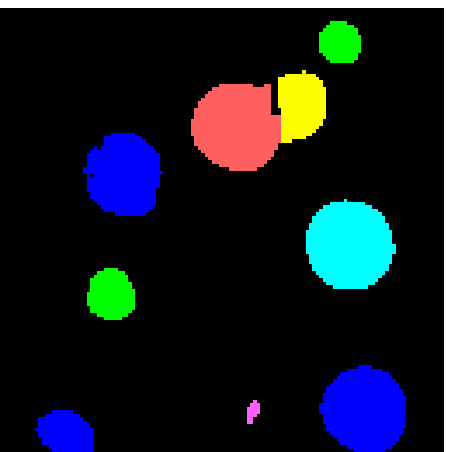,height=    2.35cm} 
	}
	\hspace{-0.3cm}
	\subfigure[]
	{
		\epsfig{figure=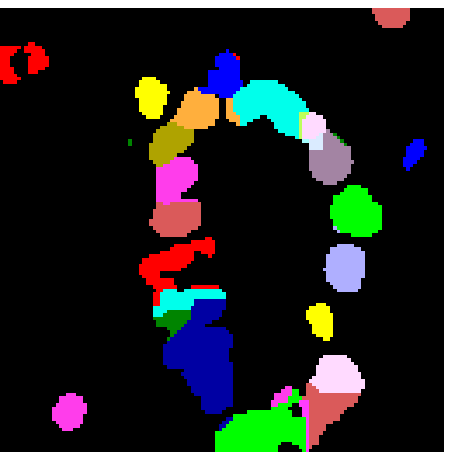,height=    2.35cm}
	}
	\hspace{-0.3cm}
	\subfigure[]
	{
		\epsfig{figure=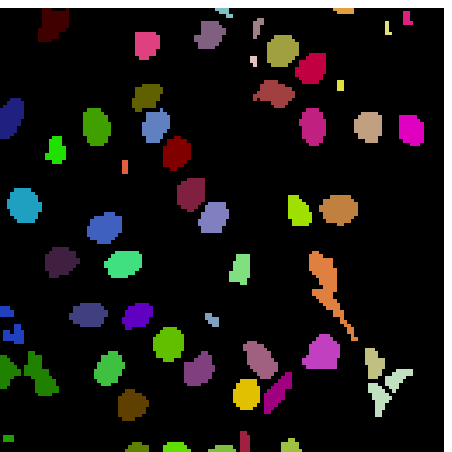,height=    2.35cm}
	}
	\hspace{-0.3cm}
	\subfigure[]
	{
		\epsfig{figure=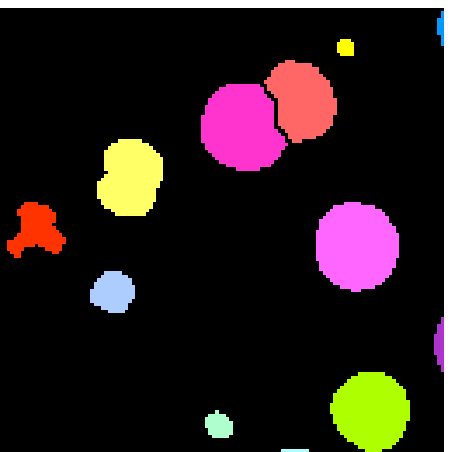,height=    2.35cm}
	}
	\hspace{-0.3cm}
	\subfigure[]
	{
		\epsfig{figure=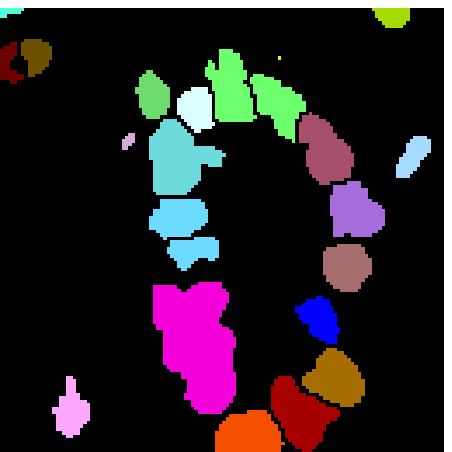,height=    2.35cm}
	}
	\caption{Examples of RCNN-SliceNet detection (first row), 3D U-Net segmentation (second row), and VTEA segmentation (third row) on original Data-I (first column), Data-II (second column) and Data-III (third column). The ground truth bounding box is in green and the detected bounding box is in red
	}
	\label{fig:rcnn_det}
	\vspace{-0.1in} 
\end{figure}

For Data-I evaluation, all models are trained on synthetic Data-I and tested on original Data-I. Similarly, for Data-II evaluation, all models are trained on synthetic Data-II and tested on original Data-II. For Data-III, we use transfer learning that all models are pre-trained on synthetic Data-III and cross validated on original Data-III. During cross validation, the $21$ volumes of original Data-III are randomly divided into 3 equal sets. We then recursively update the pre-trained model on one set and test on other two sets. For Data-IV evaluation, we test the generalizability of these models from one type of microscopy data to another. Thus, we use the model that trained on synthetic Data-III and inference it on Data-IV. 

During RCNN-SliceNet training, all slices along the $z$-direction of 50 synthetic volumes $I^{syn005}-I^{syn054}$ are used for training and validating RCNN-SliceNet. The validation set is randomly chosen 20\% from the entire training set. Scales and aspect ratios of the three anchors are set to $(64^2,128^2,256^2)$, and $(0.5,1,2)$ based on the nuclei size. 
The RCNN-SliceNet model was trained for 30 epochs with the Stochastic Gradient Descent (SGD) optimizer and an initial learning rate of 0.001 that decays 50\% every 5 epochs. The training images are 
normalized using Caffe normalization method and randomly flipped horizontally and vertically while training.
During testing, the 3-way inference was used to run RCNN-SliceNet along the $x$-direction, the $y$-direction, and the $z$-direction of a volume. Note that for Data-III and Data-IV, due to the small number of slices on the $z$-direction, we only used RCNN-SliceNet on the $z$-direction.
\subsection{Evaluation Metrics}
\label{ssec:evaluation_metrics}
The Mean Absolute Percentage Error (MAPE) is used to estimate the counting accuracy.
\begin{align}\label{eq:MAPE}
	\mathrm{MAPE}=\frac{100\%}{N}{\sum_{i=1}^{N}\left|\frac{N^{cluster}_i - N^{gt}_i }{N^{gt}_i }\right|}
\end{align}
where $N$ is the number of volumes, $N^{cluster}_i$ represents the estimated number of nuclei in $i^{th}$ volume, and $N^{gt}_i$ is the ground truth number of nuclei for $i^{th}$ volume. 
The centroid-based nuclei detection accuracy is evaluated using Average Precision (AP) and mean Average Precision (mAP) inspired by \cite{everingham2015pascal}. 
If the Euclidean distance between an estimated centroid and a ground truth centroid is less than $T_{dist}$, then we say the estimated centroid and ground truth centroid are matched. We use greedy matching which means each ground truth centroid always matches with its nearest detected centroid.
We then estimate the True Positives (number of matched estimated centroids and ground truth centroids), False Positives (number of estimated centroids that have no associated ground truth centroids matched), False Negatives (number of ground truth centroids that have no associated estimated centroids matched), and the total number of nuclei in a volume, respectively. Also, we define $\mathrm{AP}_t$ as the average precision when $T_{dist}$ is set to $t$ where $t\in\{4,5,...,8\}$, $\{6,7,...,12\}$, $\{6,7,...,10\}$, and $\{15,16,...,25\}$ for Data-I, Data-II, Data-III, and Data-IV. 
The AP is estimated by the area under the precision/recall curve \cite{everingham2015pascal} and the mAP is estimated using Equation \ref{eq:mAP}. For every detected centroid, we assume their confidences are all 1.
\begin{align}\label{eq:mAP}
	\mathrm{mAP} = \frac{1}{|T_{dist}|}\sum_{t\in T_{dist}}{\mathrm{AP}(t)}
\end{align}
where $\mathrm{AP}(t)$ is the Average Precision given a distance threshold $T_{dist}=t$.
The evaluation accuracy is shown in Table \ref{table_1}. 
\subsection{Discussion}
\label{ssec:performance_analysis}
\begin{figure*}[htb!]
	\centering
	\subfigure[]
	{
		\epsfig{figure=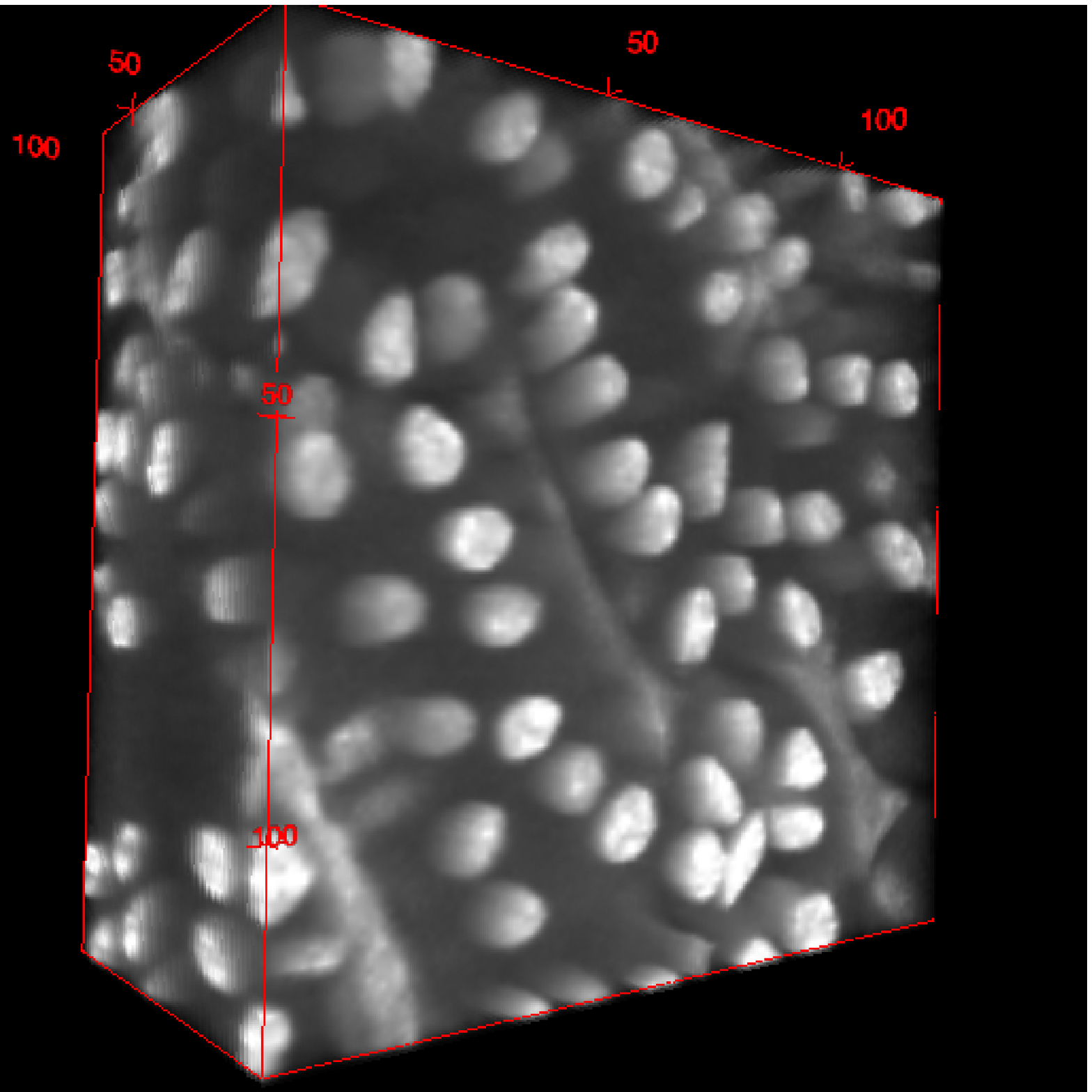,width=3.4cm}
	}
	\hspace{-0.3cm}
	\vspace{-0.3cm}
	\subfigure[]
	{
		\epsfig{figure=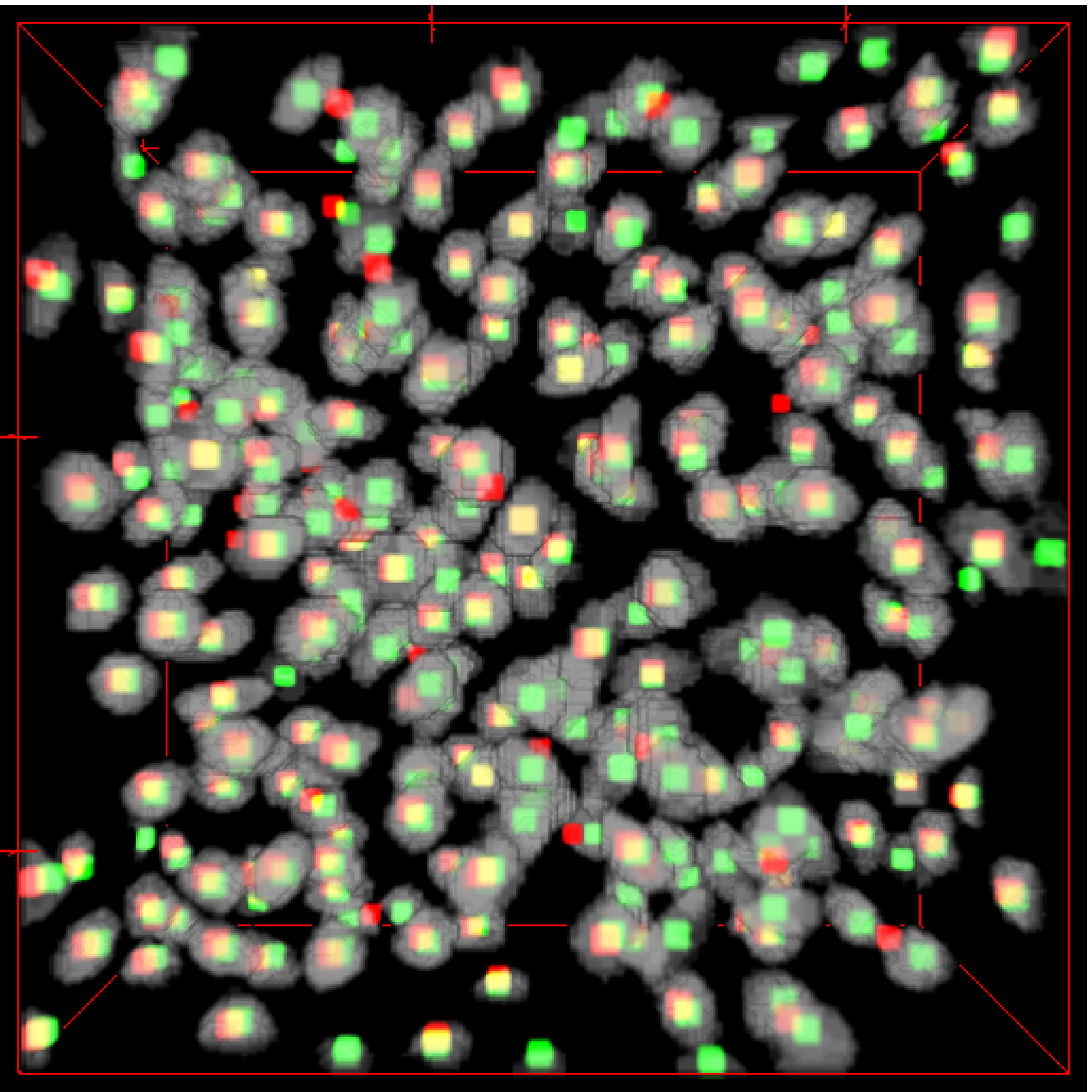,width=3.4cm}
	}
	\hspace{-0.3cm}
	\subfigure[]
	{
		\epsfig{figure=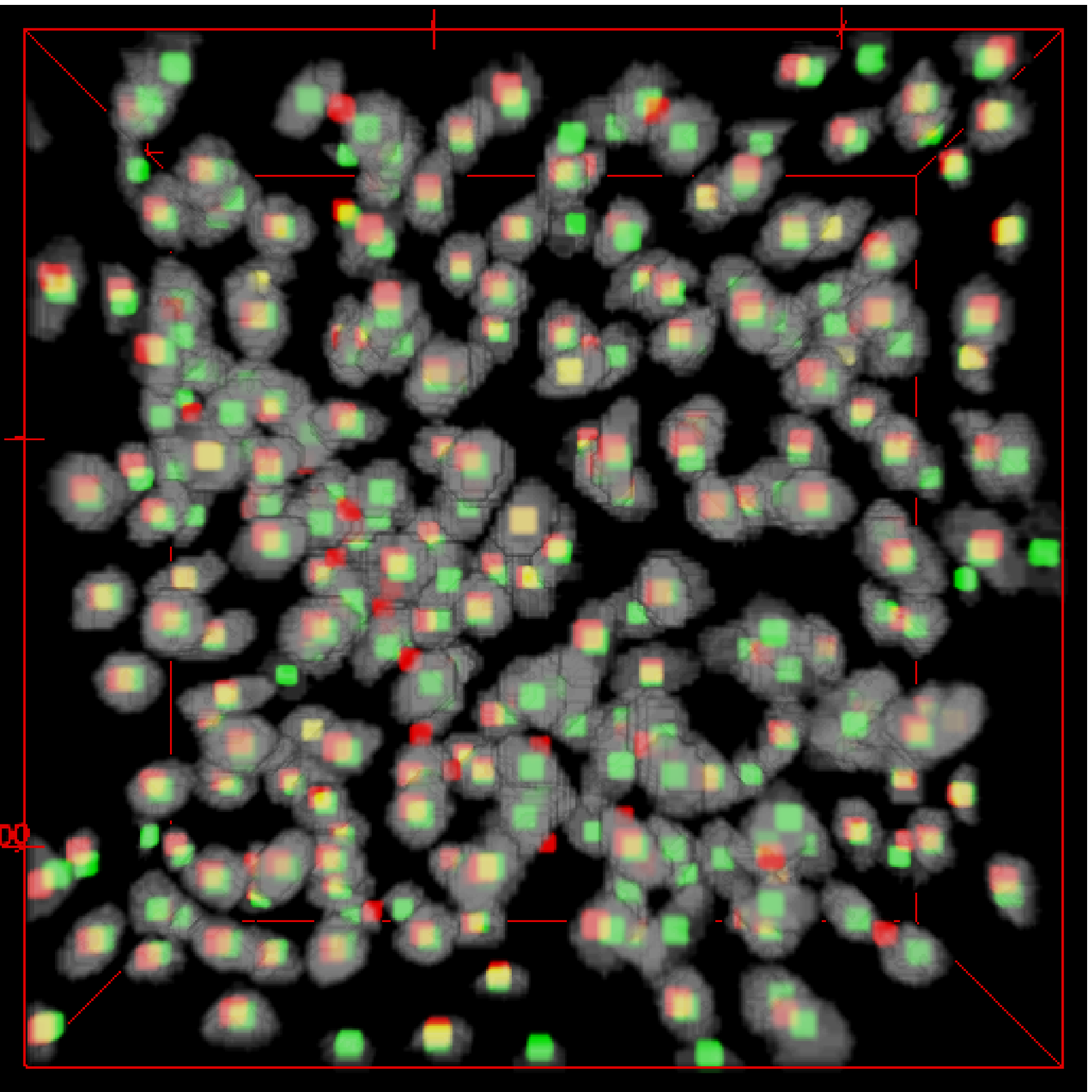,width=3.4cm}
	}
	\hspace{-0.3cm}
	\subfigure[]
	{
		\epsfig{figure=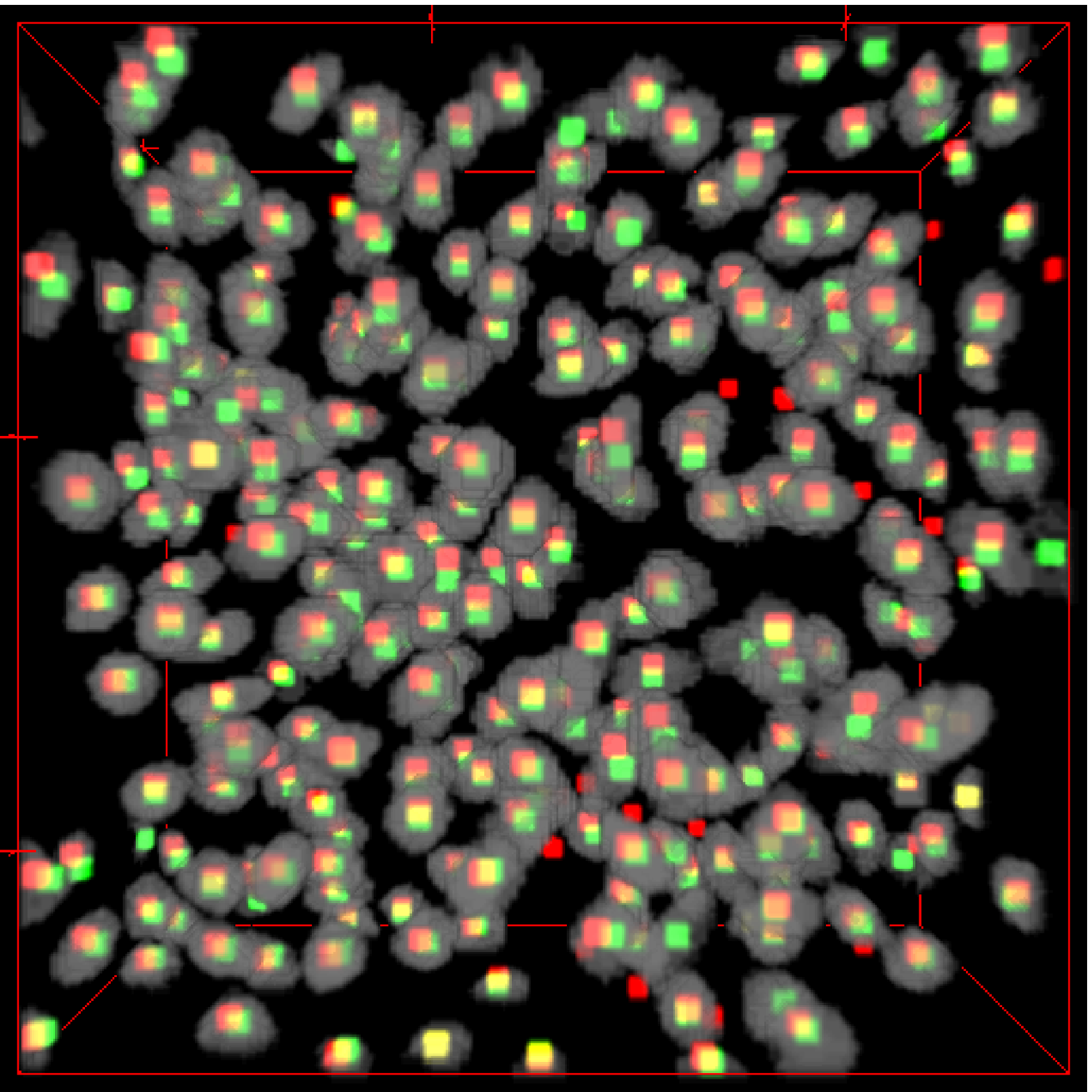,width=3.4cm}
	}
	\hspace{-0.3cm}
	\subfigure[]
	{
		\epsfig{figure=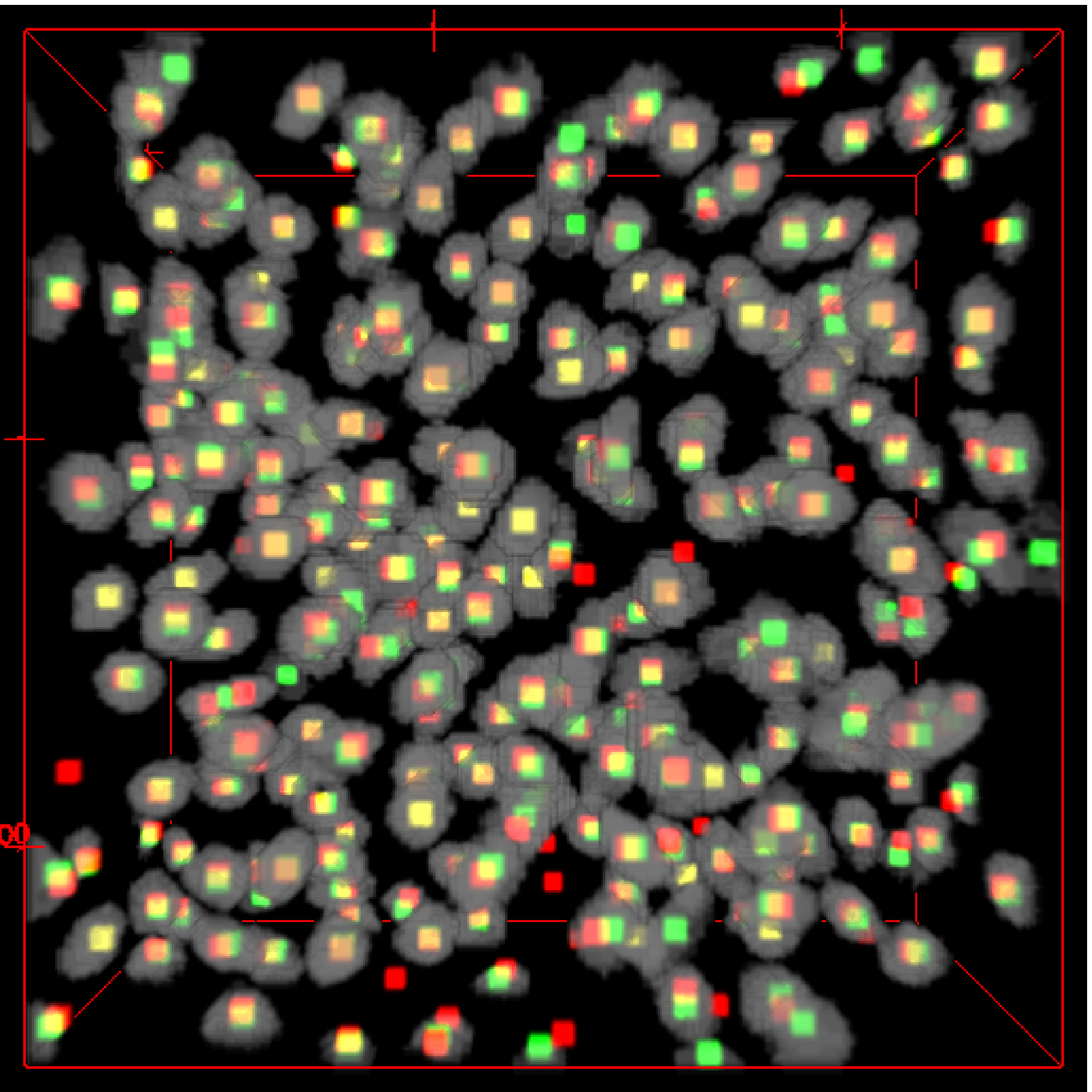,width=3.4cm}
	}
	\subfigure[]
	{
		\epsfig{figure=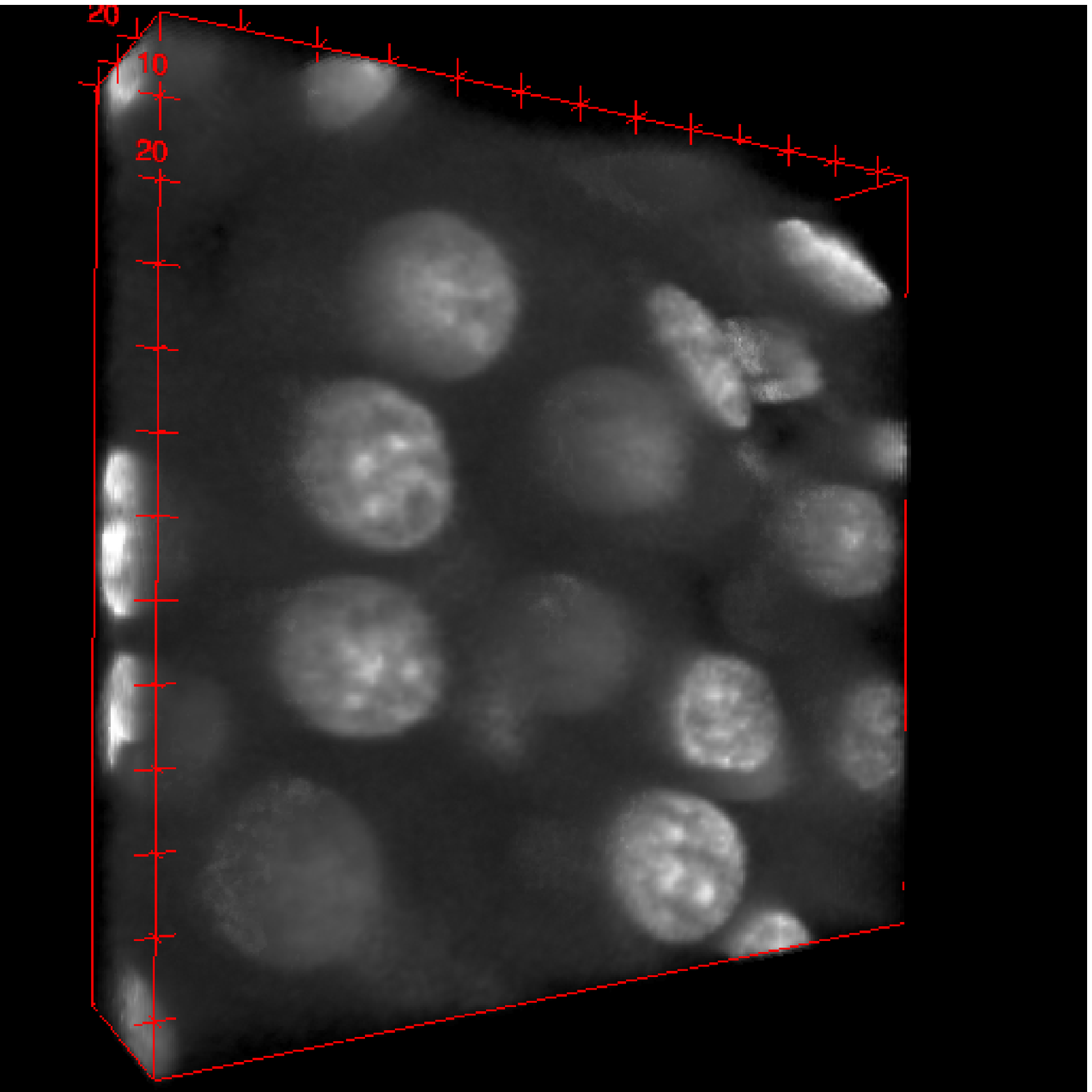,width=3.4cm}
	}
	\hspace{-0.3cm}
	\vspace{-0.1cm}
	\subfigure[]
	{
		\epsfig{figure=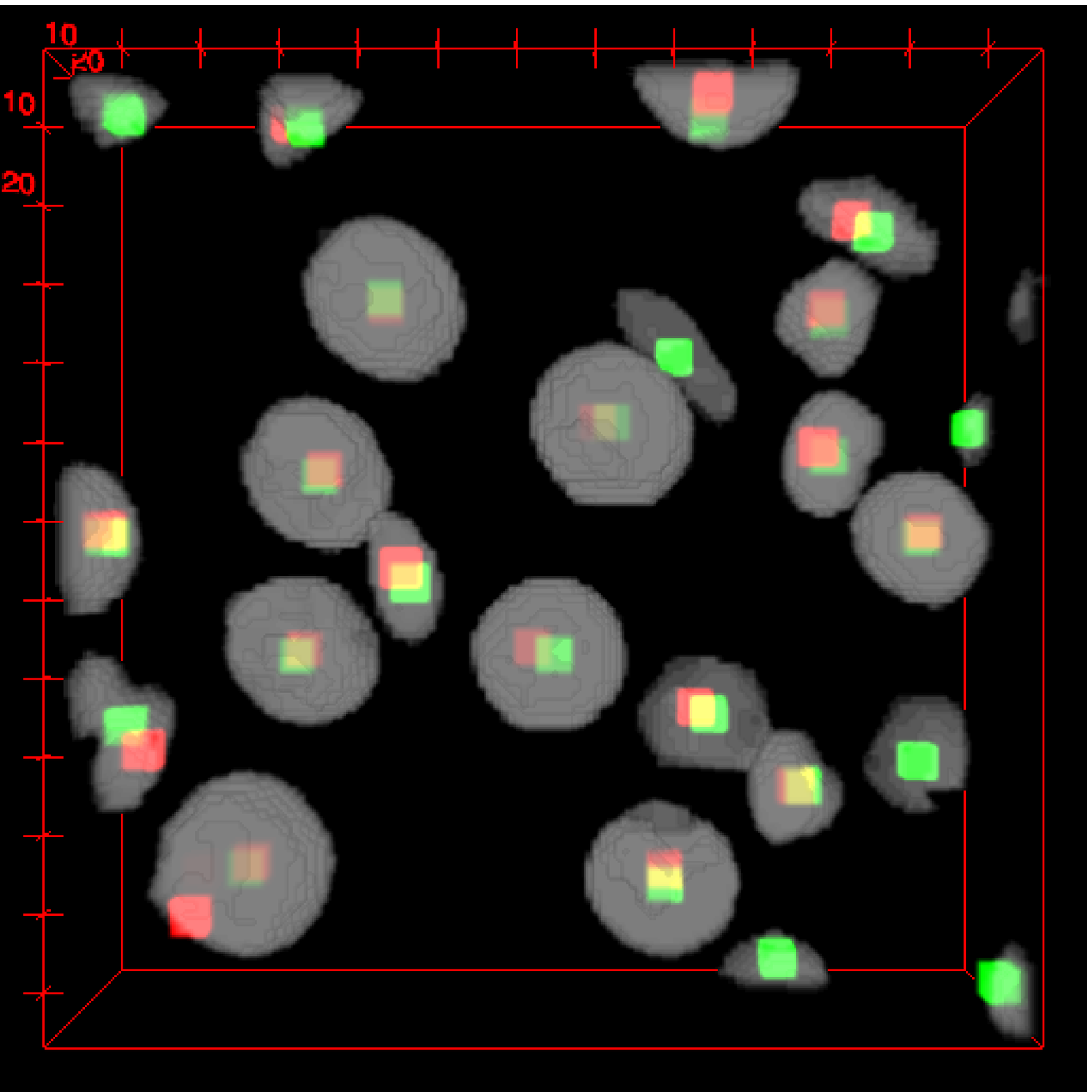,width=3.4cm}
	}
	\hspace{-0.3cm}
	\subfigure[]
	{
		\epsfig{figure=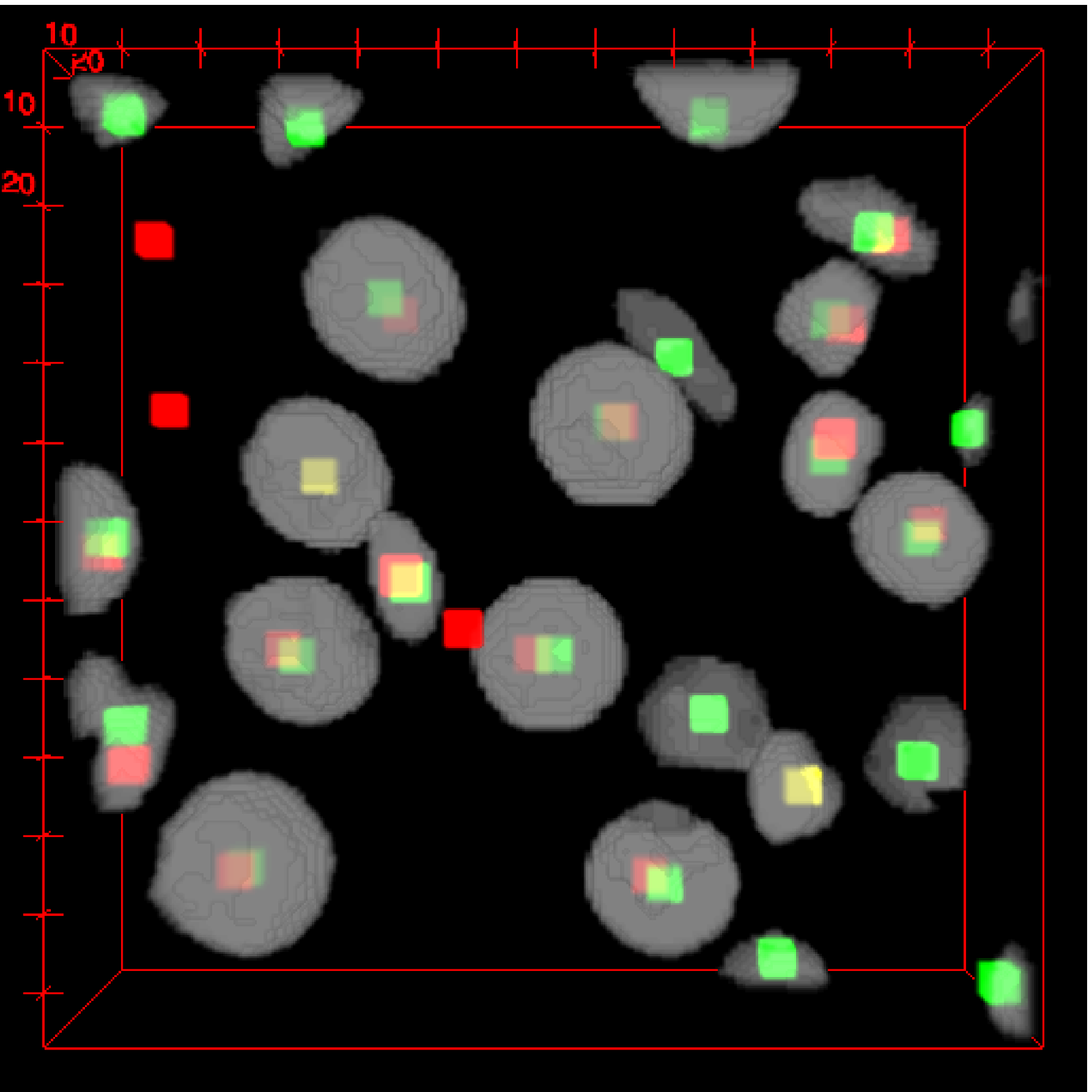,width=3.4cm}
	}
	\hspace{-0.3cm}
	\subfigure[]
	{
		\epsfig{figure=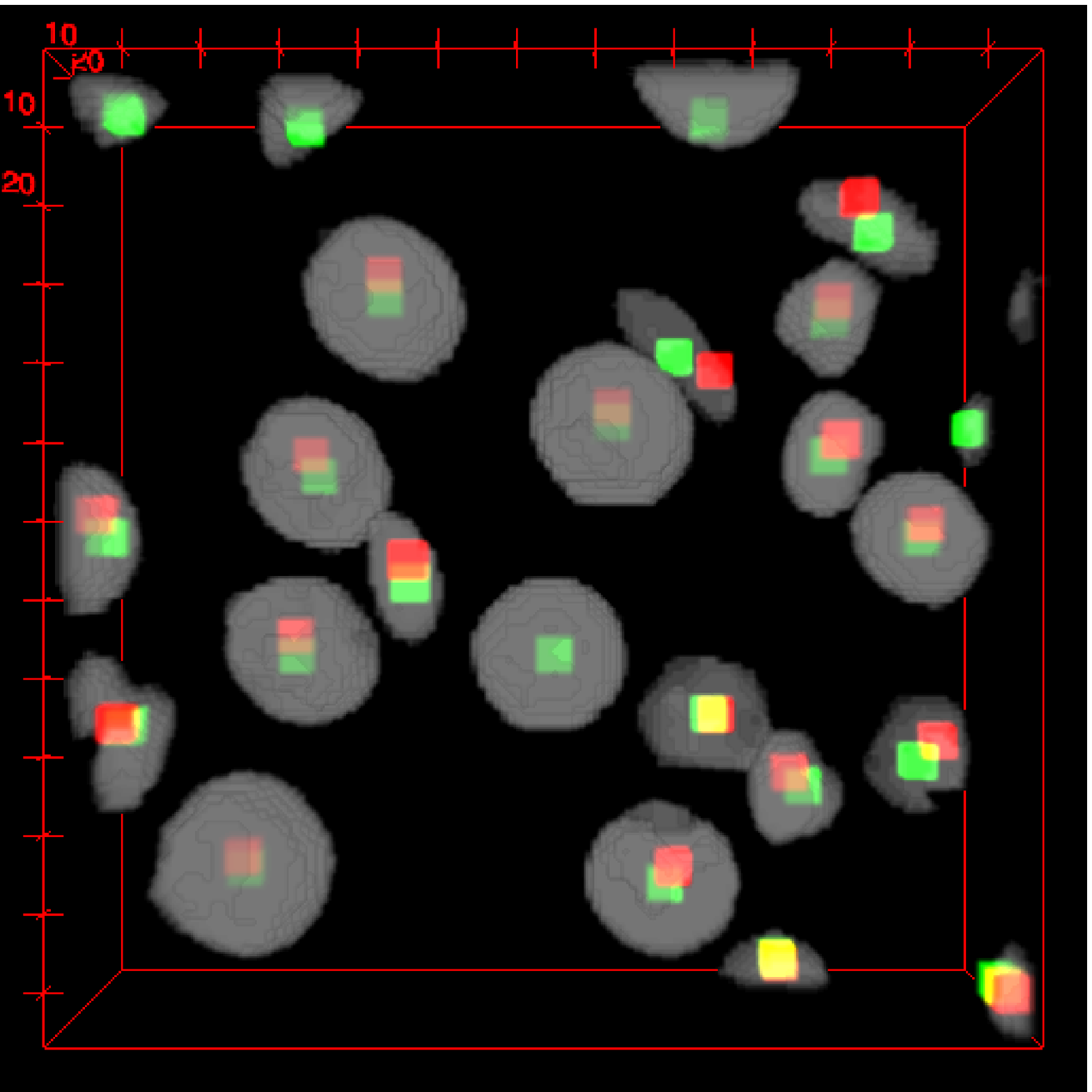,width=3.4cm}
	}
	\hspace{-0.3cm}
	\vspace{-0.2cm}
	\subfigure[]
	{
		\epsfig{figure=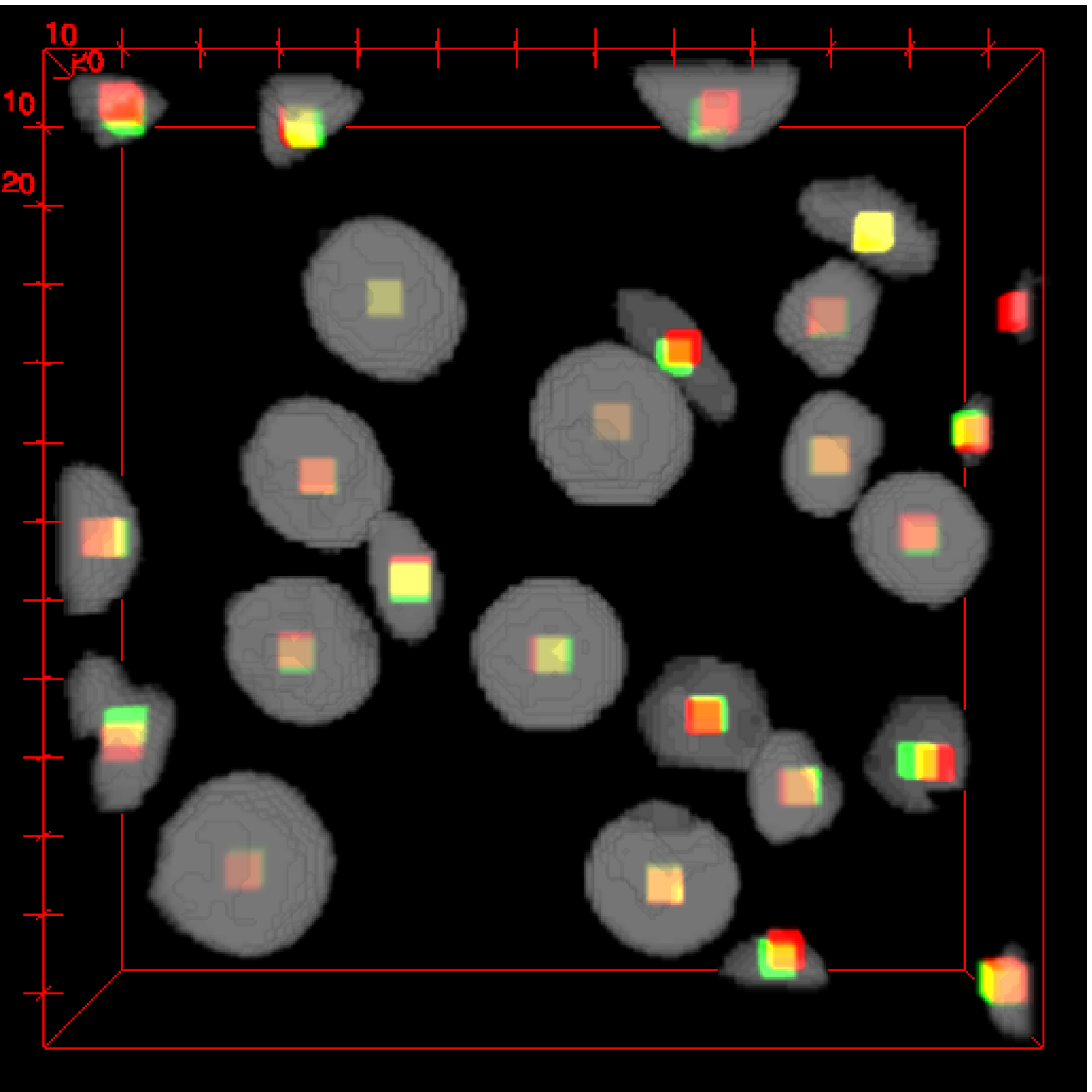,width=3.4cm}
	}
	\subfigure[]
	{
		\epsfig{figure=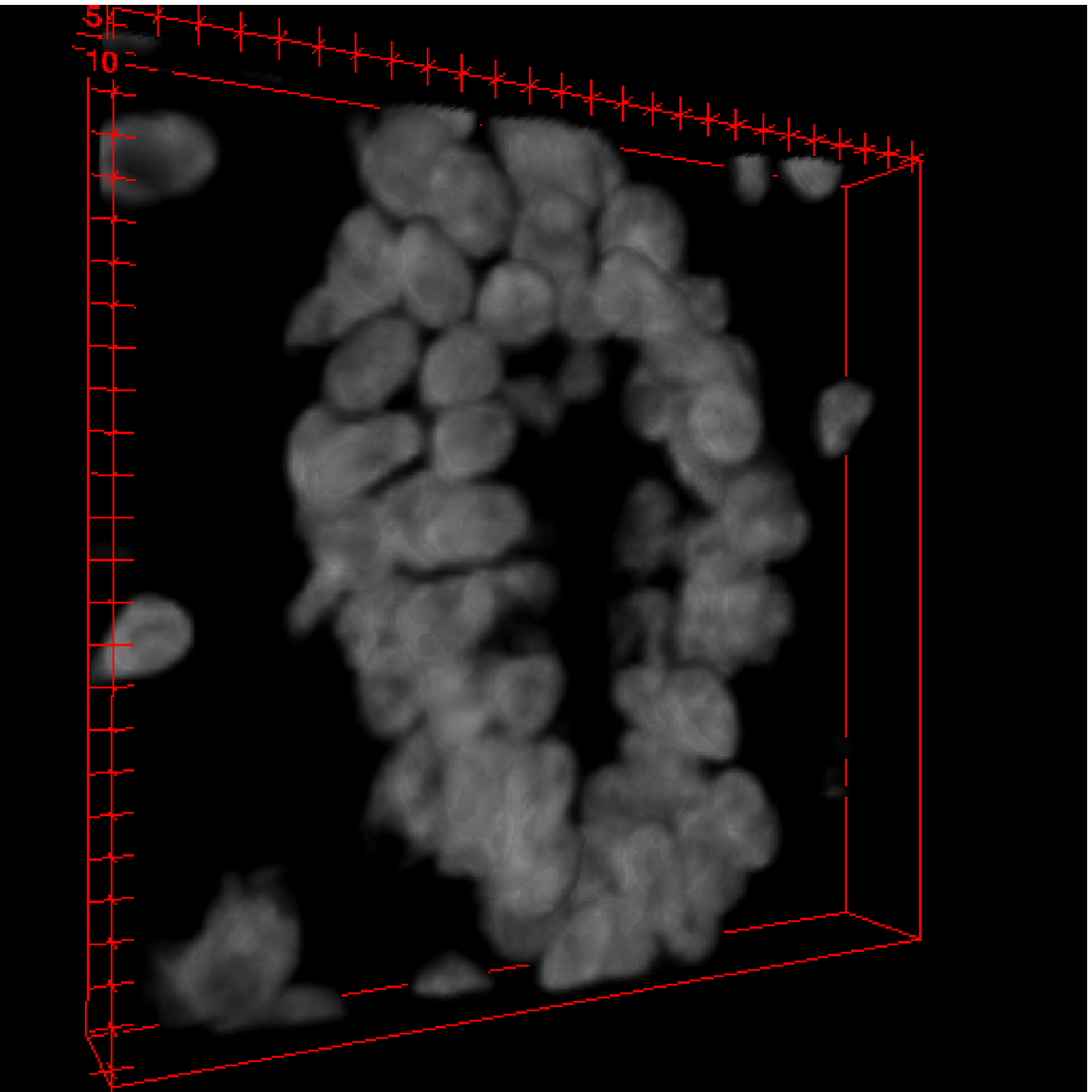,width=3.4cm}
	}
	\hspace{-0.3cm}
	\vspace{-0.1cm}
	\subfigure[]
	{
		\epsfig{figure=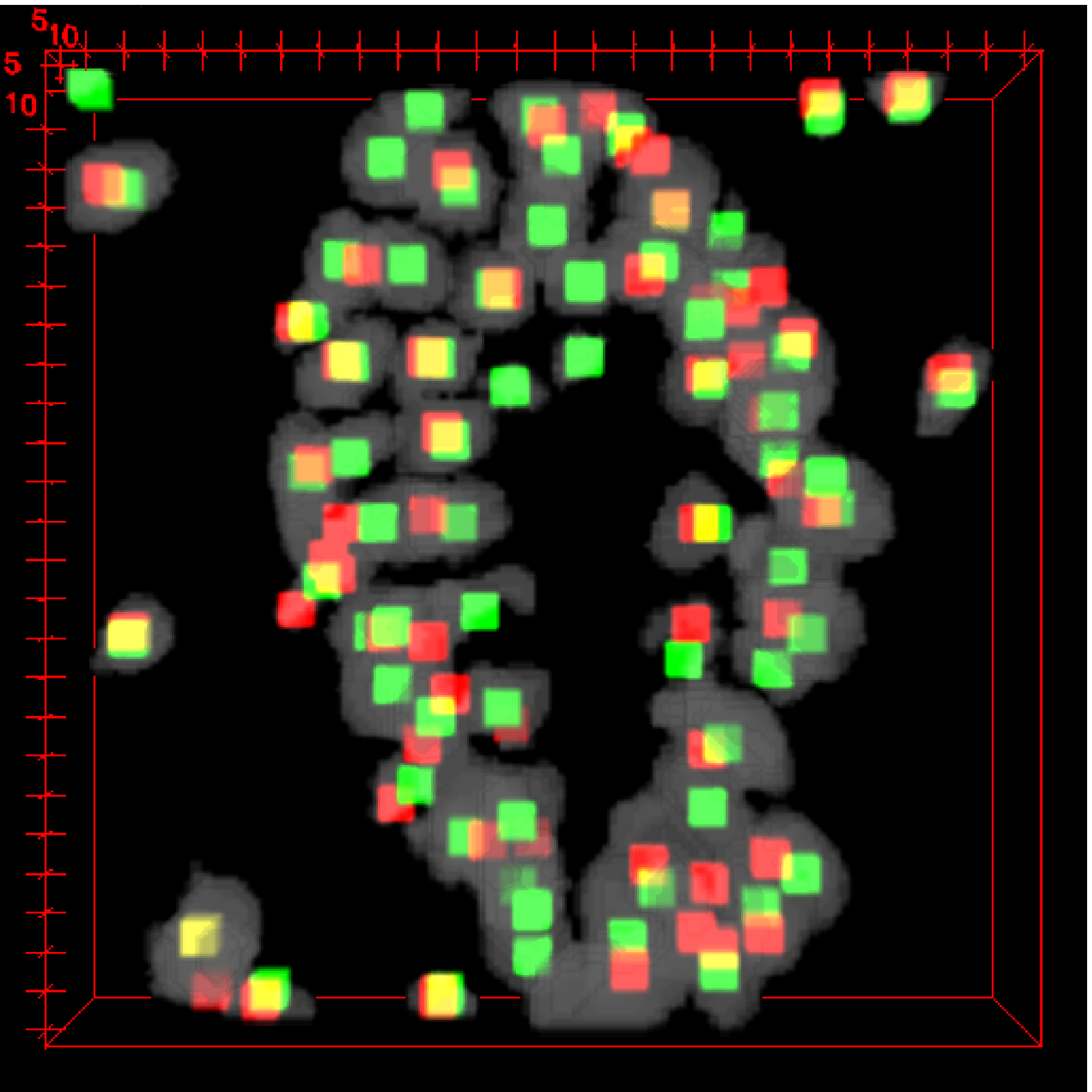,width=3.4cm}
	}
	\hspace{-0.3cm}
	\subfigure[]
	{
		\epsfig{figure=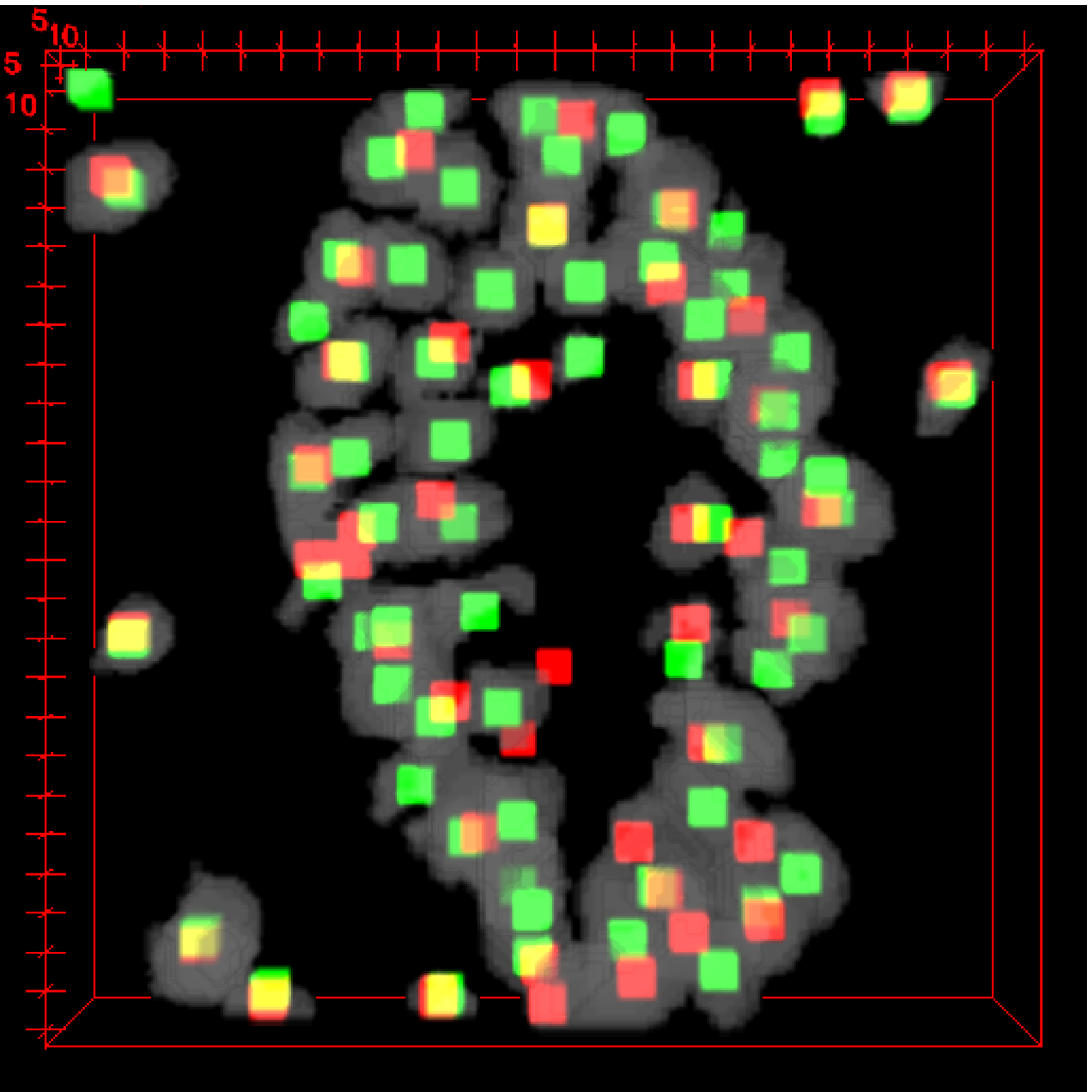,width=3.4cm}
	}
	\hspace{-0.3cm}
	\subfigure[]
	{
		\epsfig{figure=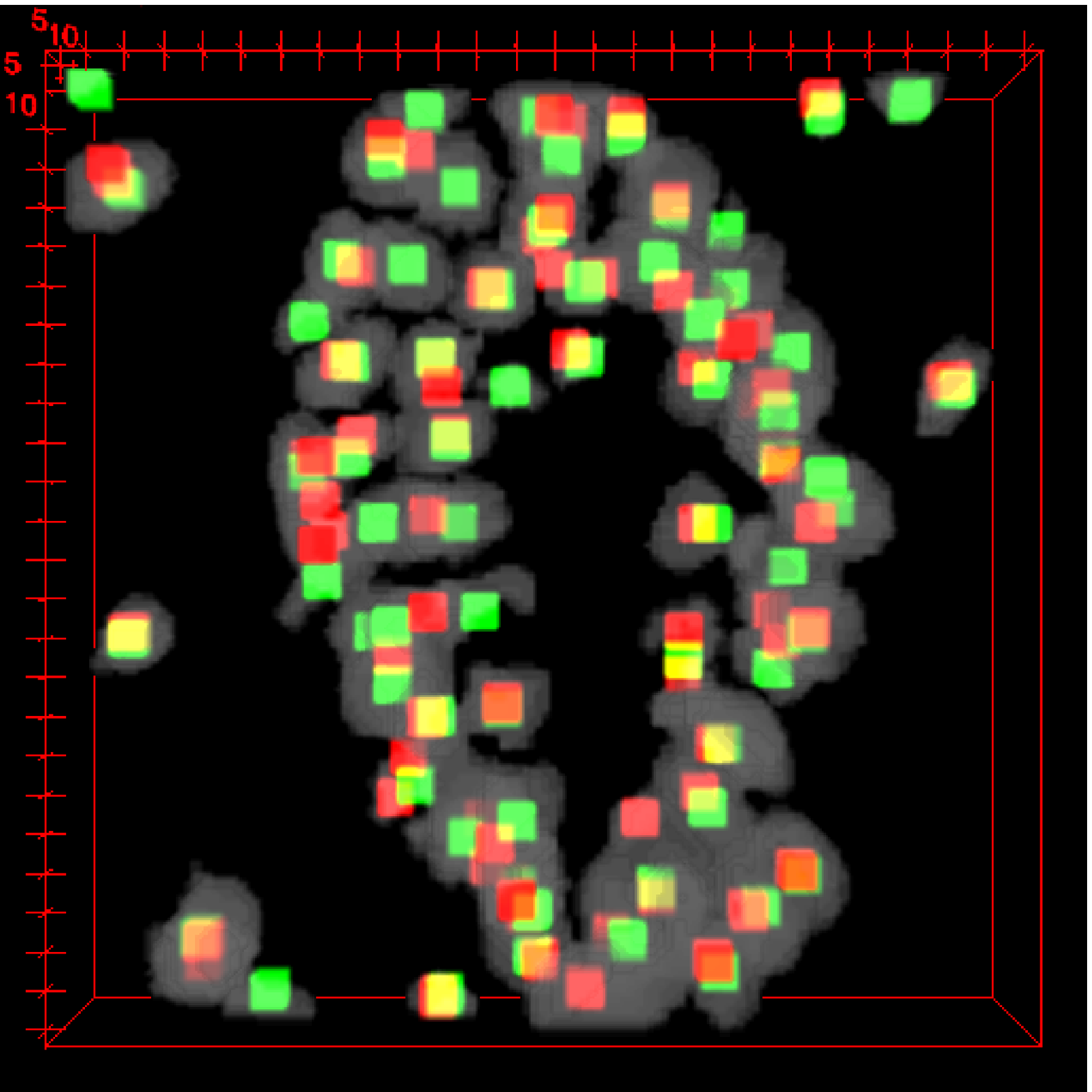,width=3.4cm}
	}
	\hspace{-0.3cm}
	\subfigure[]
	{
		\epsfig{figure=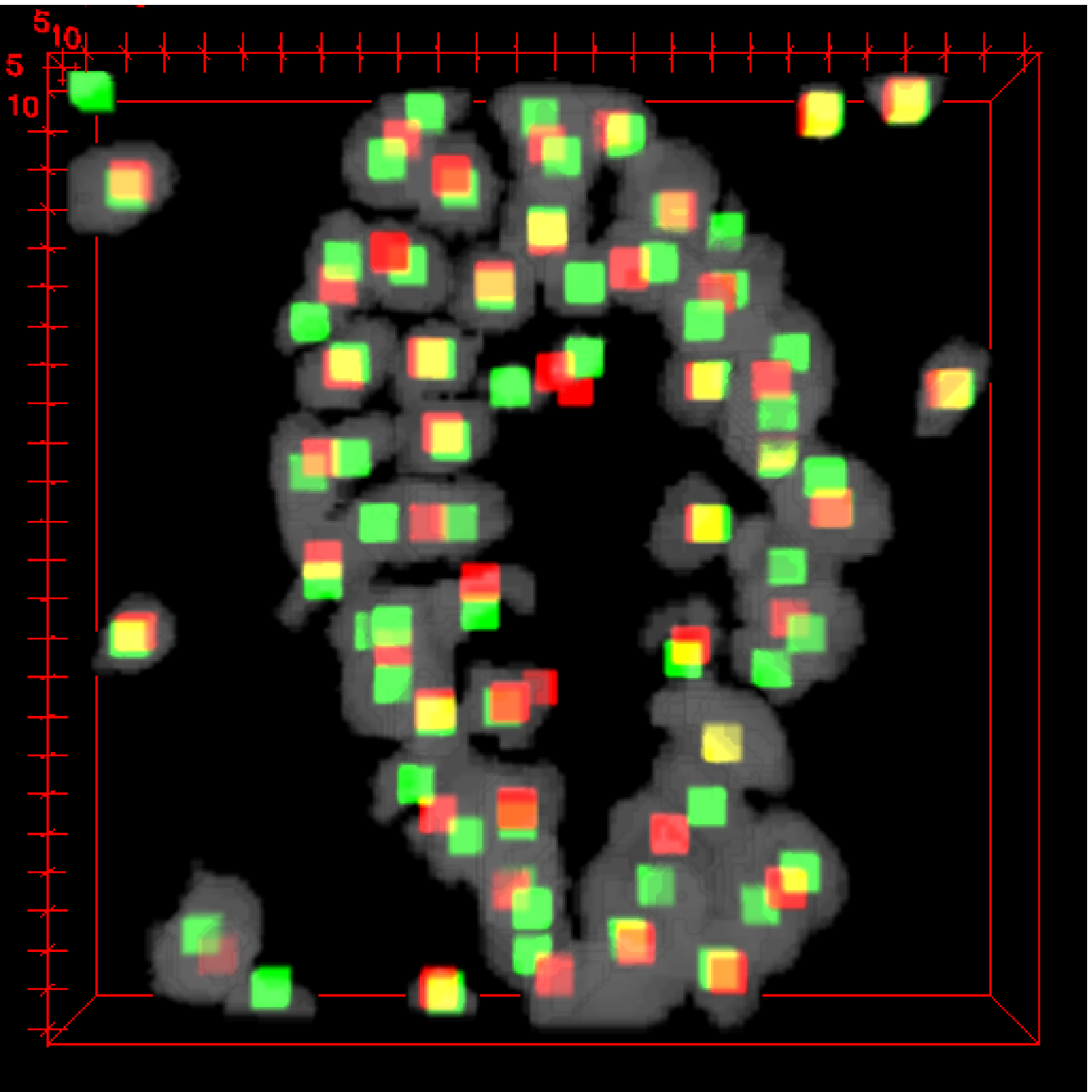,width=3.4cm}
	}
	\caption{(a), (f), (k) are example testing volumes from original microscopy Data-I, Data-II, and Data-III. The nuclei centroid estimation results for 3D U-Net (second column), V-Net (third column), DeepSynth (fourth column), and proposed method (fifth column). The red cubes are the estimated centroids, the green cubes are ground truth centroids, and the yellow is the overlay of green cubes and red cubes. The gray spheroid is the ground truth mask of the nuclei. The volumes are visualized using the ImageJ's 3D Viewer}
	\label{fig:3d_vis}
\end{figure*}
Our proposed method was compared with 3D marker-controlled watershed, which was inspired by \cite{4012364}, V-Net \cite{VNet}, 3D U-Net \cite{3DUnet},
DeepSynth \cite{deepsynth}, and other commonly used segmentation methods.
For VTEA \cite{winfree2017quantitative}, ImageJ's JACoP \cite{bolte2006guided}, Squassh \cite{Squassh}, and CellProfiler \cite{CellProfiler}, we used contrast enhancement, background subtraction, and gaussian blur to pre-process the image.
For V-Net and 3D U-Net, we optimize the results by using marker-controlled 3D watershed as the post-processing to separate touching nuclei, the markers are obtained by using conditional erosion \cite{4012364}. 
Finally, 3D connected components test was used to estimate the total number of nuclei and the nuclei centroids.

For our proposed method, we observed some outliers from the output of RCNN-SliceNet that may affect the evaluation accuracy. We use an extra 3D AHC for majority voting to remove outliers. 
As shown in Figure \ref{fig:rcnn_det}, RCNN based method can easily distinguish different nuclei and can still capture small partially included objects.
The segmentation-based methods V-Net, 3D U-Net and DeepSynth achieve good segmentation accuracy but post-processing such as watershed is required to separate different objects. 
As shown in Table \ref{table_1}, our proposed method outperforms all of the baseline methods. Table \ref{table_2} shows that our method achieved best generalization results on original Data-IV using the pre-trained model on synthetic Data-III.
Figure \ref{fig:3d_vis} shows that our method made fewer mistakes when nuclei are overlapping together and can detect the small nuclei even they are on the border of the volume.

\textbf{Advantages}\quad 
By making use of RCNN, our system can easily distinguish touching nuclei without any post-processing steps such as watershed or morphological operation. Our method can better capture small nuclei and partially included nuclei on the border (see Figure \ref{fig:3d_vis} (j)).
The slice-and-cluster strategy is robust handling the errors from RCNN-SliceNet's misdetections on some slices. The 3D AHC will treat RCNN-SliceNet's misdetection as outliers and 3-way majority voting will further align the centroid location and remove outliers. 
Our methods achieved better results than segmentation-based methods on ellipsoidal nuclei data.  

\textbf{Limitations}\quad 
Our method is under the assumption that nuclei are ellipsoidal shape. We observed the 3D AHC made more mistakes on clustering centroids for non-ellipsoidal nuclei volumes. Also, the 3D AHC requires a rough estimation of the number of nuclei in a volume to accelerate the inference time. For example, between $10$ and $100$. 
Our method is subject to over-detection problem and the error is mainly from false positive detection due to the imbalanced training samples.

\section{Conclusions}
\label{sec:conclusion}
In this paper, we presented a nuclei counting and localization method using synthetic microscopy volume generated from SpCycleGAN. We proposed a scalable framework RCNN-SliceNet using the slice-and-cluster strategy to detect each instance of nucleus on each focal plane. Our proposed system extends RCNN capability from 2D images to 3D volumes by using 3D AHC to estimate the centroid information of each nucleus. 
The experiments show that the proposed system can successfully distinguish in focus and out of focus nuclei as well as the background structures. 
Our method can accurately count the number of nuclei and estimate the nuclei centroids in a 3D microscopy volume and outperforms the rest of the methods. 
In the future, we plan to extend SpCycleGAN so it can generate different types of synthetic nuclei in one volume for training the RCNN-SliceNet, and we will improve and redesign our method to count and localize different types of nuclei in a single 3D volume.
\section{Acknowledgments}
\label{sec:acknowledgments}
This work was partially supported by a George M. O'Brien Award from the National Institutes of Health under grant NIH/NIDDK P30 DK079312 and the endowment of the Charles William Harrison Distinguished Professorship at Purdue University. 
	The authors have no conflicts of interest.
	Address all correspondence to Edward J. Delp, ace@ecn.purdue.edu
{\small
	\bibliographystyle{IEEEtran}
	\bibliography{refs.bib}
}
	
\end{document}